\documentclass{elsart}
\usepackage{amsmath}
\usepackage{graphicx}
\usepackage{amsfonts}
\usepackage{amssymb}
\newtheorem{theorem}{Theorem}

\newtheorem{lemma}[theorem]{Lemma}

\newenvironment{proof}[1][Proof]{\textbf{#1.} }{\ \rule{0.5em}{0.5em}}
\begin{document}
\begin{frontmatter}
\title{Exact form factors in integrable quantum field theories: the scaling
$Z(N)$-Ising model}
\author{Hratchya Babujian}\thanks{e-mail: babujian@physik.fu-berlin.de}
\address{Yerevan Physics Institute,
\\Alikhanian Brothers 2, Yerevan, 375036 Armenia}
\author{Angela Foerster}\thanks{e-mail: angela@if.ufrgs.br}
\address{Instituto de F\'{\i}sica da UFRGS,\\
Av. Bento Gon\c{c}alves 9500, Porto Alegre, RS - Brazil}
\author{Michael Karowski}\thanks{e-mail: karowski@physik.fu-berlin.de}
\address{Institut f\"{u}r Theoretische Physik Freie Universit\"at Berlin,\\
Arnimallee 14, 14195 Berlin, Germany}
\begin{abstract}
A general form factor formula for the scaling $Z(N)$-Ising model is
constructed. Exact expressions of all matrix elements are obtained for several
local operators. In addition, the commutation rules for order, disorder
parameters and para-Fermi fields are derived. Because of the unusual
statistics of the fields, the quantum field theory seems to be not related to
any classical Lagrangian or field equation.\\[8pt]
\end{abstract}
\end{frontmatter}

\section{Introduction}

The `form factor program' is part of the so-called `bootstrap program' for
integrable quantum field theories in 1+1-dimensions. This program
\emph{classifies} integrable quantum field theoretic models and in addition it
provides their explicit exact solutions in terms of all Wightman functions.
This means, in particular, that we do \underline{not} \emph{quantize }a
classical field theory.
In fact the quantum field theory considered in this paper is not related
(at least to our knowledge) to any classical Lagrangian or field equations
of massive particles.
The reason for this seems to be the
unusual anyonic statistics of the fields, turning this form factor
investigations even more fascinating.
The bootstrap program consists of three
main steps: First the S-matrix is calculated by means of general properties as
unitarity and crossing, the Yang-Baxter equations and the additional
assumption of `maximal analyticity'. Second, matrix elements of local
operators
\[
^{out}\left\langle \,p_{m}^{\prime},\ldots,p_{1}^{\prime}\mid\mathcal{O}%
(x)\mid p_{1},\ldots,p_{n}\,\right\rangle ^{in}\,
\]
are calculated using the 2-particle S-matrix as an input. As a third step the
Wightman functions can be obtained by inserting a complete set of intermediate
states.

The generalized form factors \cite{KW} are defined by the vacuum -
$n$-particle matrix elements
\[
\langle\,0\mid\mathcal{O}(x)\mid p_{1},\dots,p_{n}\,\rangle_{\alpha_{1}%
\dots\alpha_{n}}^{in}=e^{-ix(p_{1}+\dots+p_{n})}\,F_{\alpha_{1}\dots\alpha
_{n}}^{\mathcal{O}}(\theta_{1},\dots,\theta_{n})
\]
where the $\alpha_{i}$ denote the type (charge) and the $\theta_{i}$ are the
rapidities of the particles $\left(  p_{i}=M_{i}(\cosh\theta_{i},\sinh
\theta_{i})\right)  $. This definition is meant for $\theta_{1}>\dots
>\theta_{n}$, in the other sectors of the variables the function
$F_{\underline{\alpha}}^{\mathcal{O}}(\underline{\theta})=\,F_{\alpha_{1}%
\dots\alpha_{n}}^{\mathcal{O}}(\theta_{1},\dots,\theta_{n})$ is given by
analytic continuation with respect to the $\theta_{i}$. General matrix
elements are obtained from $F_{\underline{\alpha}}^{\mathcal{O}}%
(\underline{\theta})$ by crossing which means in particular the analytic
continuation $\theta_{i}\rightarrow\theta_{i}\pm i\pi$. Using general LSZ
assumptions and maximal analyticity in \cite{BFKZ} the following properties of
form factors have been derived:\footnote{The formulae have been proposed in
\cite{Sm} as a generalization of formulae in \cite{KW}. The formulae are
written here for the case of no backward scattering, for the general case see
\cite{BK}.}

\begin{enumerate}
\item[(o)] The form factor function $F_{\underline{\alpha}}^{\mathcal{O}%
}({\underline{\theta}})$ is meromorphic with respect to all variables
$\theta_{1},\dots,\theta_{n}$.

\item[(i)] It satisfies Watson's equations
\[
F_{\dots\alpha_{i}\alpha_{j}\dots}^{\mathcal{O}}(\dots,\theta_{i},\theta
_{j},\dots)=F_{\dots\alpha_{j}\alpha_{i}\dots}^{\mathcal{O}}(\dots,\theta
_{j},\theta_{i},\dots)\,S_{\alpha_{i}\alpha_{j}}(\theta_{ij}).
\]

\item[(ii)] The crossing relation means for the connected part (see
e.g.~\cite{BK}) of the matrix element
\begin{align*}
_{\bar{\alpha}_{1}}\langle\,p_{1}\mid\mathcal{O}(0)\mid p_{2},\dots
,p_{n}\,\rangle_{\alpha_{2}\dots\alpha_{n}}^{in,conn.}  &  =\sigma
^{\mathcal{O}}(\alpha_{1})\,F_{\alpha_{1}\alpha_{2}\dots\alpha_{n}%
}^{\mathcal{O}}(\theta_{1}+i\pi,\theta_{2},\dots,\theta_{n})\\
&  =F_{\alpha_{2}\dots\alpha_{n}\alpha_{1}}^{\mathcal{O}}(\theta_{2}%
,\dots,\theta_{n},\theta_{1}-i\pi)
\end{align*}
where $\sigma^{\mathcal{O}}(\alpha)$ is the statistics factor of the operator
$\mathcal{O}$ with respect to the particle $\alpha$.

\item[(iii)] The function $F_{\underline{\alpha}}^{\mathcal{O}}({\underline
{\theta}})$ has poles determined by one-particle states in each sub-channel.
In particular, if $\alpha_{1}$ is the antiparticle of $\alpha_{2}$, it has the
so-called annihilation pole at $\theta_{12}=i\pi$ which implies the recursion
formula
\begin{multline*}
\operatorname*{Res}_{\theta_{12}=i\pi}F_{\underline{\alpha}}^{\mathcal{O}%
}(\theta_{1},\dots,\theta_{n})=2i\,\mathbf{C}_{\alpha_{1}\alpha_{2}}%
F_{\alpha_{3}\dots}^{\mathcal{O}}(\theta_{3},\dots,\theta_{n})\\
\times\left(  \mathbf{1}-\sigma^{\mathcal{O}}(\alpha_{2})S_{\alpha_{1}%
\alpha_{n}}(\theta_{2n})\dots S_{\alpha_{2}\alpha_{3}}(\theta_{23})\right)
\,.
\end{multline*}

\item[(iv)] Bound state form factors yield another recursion formula
\[
\operatorname*{Res}_{\theta_{12}=i\eta}F_{\alpha\beta\dots}^{\mathcal{O}%
}(\theta_{1},\theta_{2},\theta_{3},\dots)=\sqrt{2}F_{\gamma\dots}%
^{\mathcal{O}}(\theta_{(12)},\theta_{3},\dots)\,\Gamma_{\alpha\beta}^{\gamma}%
\]
if $i\eta$ is the position of the bound state pole. The bound state
intertwiner $\Gamma_{\alpha\beta}^{\gamma}$ (see e.g. \cite{K1,BK}) is defined
by
\[
i\operatorname*{Res}_{\theta=i\eta}S_{\alpha\beta}(\theta)=\Gamma_{\gamma
}^{\beta\alpha}\Gamma_{\alpha\beta}^{\gamma}%
\]

\item[(v)] Since we are dealing with relativistic quantum field theories
Lorentz covariance in the form
\[
F_{\underline{\alpha}}^{\mathcal{O}}(\theta_{1},\dots,\theta_{n})=e^{s\mu
}\,F_{\underline{\alpha}}^{\mathcal{O}}(\theta_{1}+\mu,\dots,\theta_{n}+\mu)
\]
holds if the local operator transforms as $\mathcal{O}\rightarrow e^{s\mu
}\mathcal{O}$ where $s$ is the \textquotedblleft spin\textquotedblright\ of
$\mathcal{O}$.

\end{enumerate}

Note that consistency of (ii), (iii) and (v) imply a relation of spin and
statistics $\sigma^{\mathcal{O}}(\alpha)=e^{-2\pi is}$ and also $\sigma
^{\mathcal{O}}(\alpha)=1/\sigma^{\mathcal{O}}(\bar{\alpha})$ where
$\bar{\alpha}$ is the anti-particle of $\alpha$, which has the same charge as
$\mathcal{O}$. All solutions of the form factor equations (i) - (v) should
provide the matrix elements of all fields in an integrable quantum filed
theory with a given S-matrix.

Generalized form factors are of the form \cite{KW}%
\begin{equation}
F_{\underline{\alpha}}^{\mathcal{O}}(\underline{\theta})=K_{\underline{\alpha
}}^{\mathcal{O}}(\underline{\theta})\prod_{1\leq i<j\leq n}F(\theta
_{ij})\,,\quad\left(  \theta_{ij}=\theta_{i}-\theta_{j}\right)  \label{4}%
\end{equation}
where $F(\theta)$ is the 'minimal' form factor function. It is the solution of
Watsons equation \cite{Wa} and the crossing relation for $n=2$
\begin{equation}%
\begin{array}
[c]{c}%
F(\theta)=F(-\theta)S(\theta)\\
F(i\pi-\theta)=F(i\pi+\theta)
\end{array}
\label{3}%
\end{equation}
with no poles and zeros in the physical strip $0<\operatorname{Im}\theta
\leq\pi$ (and a simple zero at $\theta=0$). In \cite{BK} a general integral
representation for the K-function $K_{\underline{\alpha}}^{\mathcal{O}%
}(\underline{\theta})$ in terms of the \emph{off-shell Bethe Ansatz}
\cite{B,B1} has been presented, which transforms the complicated equations
(i)-(v) for the form factors to simple ones for the \emph{p-functions} (see
\cite{BK} and (\ref{p}) below)
\begin{equation}
K_{\underline{\alpha}}^{\mathcal{O}}(\underline{\theta})=\sum_{m}c_{nm}%
\int_{\mathcal{C}_{\underline{\theta}}}dz_{1}\cdots\int_{\mathcal{C}%
_{\underline{\theta}}}dz_{m}\,h(\underline{\theta},\underline{z}%
)\,p^{\mathcal{O}}(\underline{\theta},\underline{z})\,\Psi_{\underline{\alpha
}}(\underline{\theta},\underline{z})\ . \label{K}%
\end{equation}
The symbols $\mathcal{C}_{\underline{\theta}}$ denote specific contours in the
complex $z_{i}$-planes. The function $h(\underline{\theta},\underline{z}) $ is
scalar and encodes only data from the scattering matrix. The function
$p^{\mathcal{O}}(\underline{\theta},\underline{z})$ on the other hand depends
on the explicit nature of the local operator $\mathcal{O}$. We discuss these
objects in more detail below. For the case of a diagonal S-matrix, as in this
paper, the off-shell Bethe Ansatz vector $\Psi_{\underline{\alpha}}%
(\underline{\theta},\underline{z})$ is trivial. The K-function $K_{\underline
{\alpha}}^{\mathcal{O}}(\underline{\theta})$ is an meromorphic function and
has the 'physical poles' in $0<\operatorname{Im}\theta_{ij}\leq\pi$
corresponding to the form factor properties (iii) and (iv). It turns out that
for the examples we consider in this paper there is only one term in the sum
of (\ref{K}).

In this paper we will focus on the determination of the form factors of the
scaling $Z(N)$-Ising quantum field theory in 1+1 dimensions. An Euclidean
field theory is obtained as the scaling limit of a classical statistical
lattice model in 2-dimensions given by the partition function
\[
Z=\sum_{\left\{  \sigma\right\}  }\exp\left(  -\frac{1}{kT}\sum_{\left\langle
ij\right\rangle }E(\sigma_{i},\sigma_{j})\right)  \,;\;\sigma_{i}\in\left\{
1,\omega,\dots,\omega^{N-1}\right\}  ,\;\omega=e^{2\pi i/N}%
\]
as a generalization of the Ising model. It was conjectured by K\"{o}berle and
Swieca \cite{KS} that there exists a $Z(N)$-invariant interaction
$E(\sigma_{i},\sigma_{j})$ such that the resulting massive quantum field
theory is integrable. In particular for $N=2$ the scaling $Z(2)$-Ising model
is the well investigated model \cite{Bariev,MTW,SJM1,BKW} which is equivalent
to a massive free Dirac field theory. In this paper we investigate the general
$Z(N)$-model. It has also been discussed as a deformation \cite{Za,Fa} of a
conformal $Z(N)$ para-fermi field theory \cite{FZ}. The $Z(N)$-Ising model in
the scaling limit possesses $N-1$ types of particles: $\alpha=1,\dots,N-1$ of
charge $\alpha$, mass $M_{\alpha}=M\sin\pi\frac{\alpha}{N}$ and $\bar{\alpha
}=N-\alpha$ is the antiparticle of $\alpha$. The n-particle S-matrix
factorizes in terms of two-particle ones since the model is integrable. The
two-particle S-matrix for the $Z(N)$-Ising model has been proposed by
K\"{o}berle and Swieca \cite{KS}. The scattering of two particles of type $1$
is
\begin{equation}
S(\theta)=\frac{\sinh\frac{1}{2}(\theta+\frac{2\pi i}{N})}{\sinh\frac{1}%
{2}(\theta-\frac{2\pi i}{N})}\,. \label{1}%
\end{equation}
This S-matrix is consistent with the picture that the bound state of $N-1$
particles of type $1$ is the anti-particle of $1$. This will be essential also
for the construction of form factors below. We construct generalized form
factors of an operator $\mathcal{O}(x)$ and $n$ particles of type $1$ and for
simplicity we write $F_{n}^{\mathcal{O}}(\underline{\theta})=F_{1\dots
1}^{\mathcal{O}}(\underline{\theta})$. Note that all further matrix elements
with different particle states of the field operator $\mathcal{O}(x)$ are
obtained by the crossing formula (ii) and the bound state formula (iv). As an
application of this form factor approach we compute the commutation relations
of fields. In particular, we consider the fields $\psi_{Q\tilde{Q}%
}(x)\,,\;(Q,\tilde{Q}=0,\dots,N-1)$ with charge $Q$ and `dual charge'
$\tilde{Q}$. There are in particular the order parameters $\sigma_{Q}%
(x)=\psi_{Q0}(x)$, the disorder parameters $\mu_{\tilde{Q}}(x)=\psi
_{0\tilde{Q}}(x)$ and the para-fermi fields $\psi_{Q}(x)=\psi_{QQ}(x)$. We
show that they satisfy the space like commutation rules:
\begin{align}
\sigma_{Q}(x)\sigma_{Q^{\prime}}(y)  &  =\sigma_{Q^{\prime}}(y)\sigma
_{Q}(x)\nonumber\\
\mu_{\tilde{Q}}(x)\mu_{\tilde{Q}^{\prime}}(y)  &  =\mu_{\tilde{Q}^{\prime}%
}(y)\mu_{\tilde{Q}}(x)\nonumber\\
\mu_{\tilde{Q}}(x)\sigma_{Q}(y)  &  =\sigma_{Q}(y)\mu_{\tilde{Q}}%
(x)e^{\theta(x^{1}-y^{1})2\pi iQ\tilde{Q}/N}\label{cr}\\
\psi_{Q}(x)\psi_{Q}(y)  &  =\psi_{Q}(y)\psi_{Q}(x)e^{\epsilon(x^{1}-y^{1})2\pi
iQ^{2}/N}\,.\nonumber
\end{align}
It turns out that the nontrivial statistics factors $\sigma^{\mathcal{O}%
}(\alpha)$ in the form factor equations (ii) and (iii) lead the nontrivial
order- disorder and the anyonic statistics of the fields.

The form factor bootstrap program has been applied in \cite{BKW} to the $Z(2)
$-model. Form factors for the $Z(3)$-model were investigated by one of the
present authors in \cite{K3}. There the form factors of the order parameter
$\sigma_{1}$ were proposed up to four particles. Kirilov and Smirnov
\cite{KiSm} proposed all form factors of the $Z(3)$-model in terms of
determinants.
Some related work can be found in \cite{DC}.
For general $N$ form factors for charge-less states (n particles
of type 1 and n particles of type $N-1$) were calculated in \cite{JKOPS}. In
the present paper we present for the scaling $Z(N)$-Ising model integral
representations for all matrix elements of several field operators.

Recently, there has been a renewed interest in the form factors program in
connection to condensed matter physics \cite{EK,Ts,GNT} and atomic physics
\cite{LZMGo}. In particular, applications to Mott insulators and carbon
nanotubes as well as in the field of Bose-Einstein condensates of ultra-cold
atomic gases have been discussed and in some instances correlation functions
have been computed.

The paper is organized as follows: In Section 2 we construct the general form
factor formula for the simplest $N=2$ case, which corresponds to the
well-known scaling $Z(2)$ - Ising model, and show that the known results can
be reproduced by our new general approach. In Section 3 we construct the
general form factor formula for the $Z(3)$ case, which is more complex due to
the presence of bound states, and discuss several explicit examples. We extend
these results in Section 4, where the general form factors for $Z(N)$ are
constructed and discussed in detail. Section 5 contains the derivation of the
commutation rules of the fields and some applications of this formalism are
presented. Some results of the present article have been published previously
\cite{BK04} without proves.

\section{Z(2)-form factors}

To make our method more transparent and with the hope that our construction
will help to calculate form factors for all primary and descendant fields, we
start with the simplest case $N=2$, which corresponds to the well known Ising
model in the scaling limit. This model, already investigated in
\cite{Bariev,MTW,BKW,JKOPS}, is equivalent to a massive free Dirac field
theory. The model possesses one particle with mass $M$ and the 2-particle
S-matrix is $S=-1$. In \cite{BKW} the form factor approach has been applied to
this case with the result for the order parameter field $\sigma(x)$
\begin{equation}
F_{n}^{\sigma}(\underline{\theta})=\left(  2i\right)  ^{(n-1)/2}\prod_{1\leq
i<j\leq n}\tanh\tfrac{1}{2}\theta_{ij} \label{bkw}%
\end{equation}
for $n$ odd. It is easy to see that this expression satisfies the form factor
equations (i) - (iii) with statistics factor $\sigma^{\sigma}=1$. For the
$Z(2)$ case in this section we skip the proof that the functions obtained by
our general formula satisfy the form factor equations (i) - (v), the reader
may easily reduce the proofs for the $Z(3)$ and the general $Z(N)$ case of the
following sections to this simple one. Rather, we present the results for
several fields, in particular, we will show that our general formula
reproduces the known results.


\subsection{The general formula for n-particle form factors}

We propose the $n$-particle form factors of an operator $\mathcal{O}(x)$ as
given by (\ref{4}) with the minimal form factor function%
\begin{equation}
F(\theta)=\sinh\tfrac{1}{2}\theta\label{F2}%
\end{equation}
which is the minimal solution of Watson's equations and crossing (\ref{3}) for
$S=-1$. The K-function is given by our general formula (\ref{K}) where the
Bethe Ansatz vector is trivial (because there is no backward scattering) and
the sum consists only of one term%
\begin{equation}
K_{n}^{\mathcal{O}}(\underline{\theta})=N_{n}I_{nm}(\underline{\theta
},p^{\mathcal{O}})\,.\label{K2}%
\end{equation}
The \emph{fundamental building blocks} of form factors are%
\begin{align}
I_{nm}(\underline{\theta},p^{\mathcal{O}}) &  =\frac{1}{m!}\int_{\mathcal{C}%
_{\underline{\theta}}}\frac{dz_{1}}{R}\cdots\int_{\mathcal{C}_{\underline
{\theta}}}\frac{dz_{m}}{R}\,h(\underline{\theta},\underline{z}%
)\,p^{\mathcal{O}}(\underline{\theta},\underline{z})\label{Inm2}\\
h(\underline{\theta},\underline{z}) &  =\prod_{i=1}^{n}\prod_{j=1}^{m}%
\phi(z_{j}-\theta_{i})\prod_{1\leq i<j\leq m}\tau(z_{i}-z_{j})\,.\label{h2}%
\end{align}
The h-function does not depend on the operator but only on the S-matrix of the
model, whereas the p-function depends on the operator. Both are analytic
functions of $\theta_{i}\ (i=1,\dots,n)$ and $z_{j}\ (j=1,\dots,m)$ and are
symmetric under $\theta_{i}\leftrightarrow\theta_{j}$ and $z_{i}%
\leftrightarrow z_{j}$. For all integration variables $z_{j}$ the integration
contours $\mathcal{C}_{\underline{\theta}}=\sum\mathcal{C}_{\theta_{i}}$
enclose clock wise oriented the points $z_{j}=\theta_{i}\,(i=1,\dots,n)$. This
means we assume that $\phi(z)$ has a pole at $z=0$ such that $R=\int
_{\mathcal{C}_{\theta}}dz\phi(z-\theta)$. The functions $\phi\left(  z\right)
$ and $\tau(z)$ are given in terms of the minimal form factor function as
\begin{align}
\phi(z) &  =\frac{1}{F(\theta)F(\theta+i\pi)}=\frac{-2i}{\sinh z}%
\label{phi2}\\
\tau(z) &  =\frac{1}{\phi(z)\phi(-z)}=\tfrac{1}{4}\sinh^{2}z\,.\label{tau}%
\end{align}
The following properties of the p-functions guarantee that the form factors
satisfy (i) - (iii)%
\[%
\begin{array}
[c]{lll}%
\text{(i'}_{2}\text{)} &  & p_{nm}^{\mathcal{O}}(\underline{\theta}%
,\underline{z})~\text{is symmetric under }\theta_{i}\leftrightarrow\theta
_{j}\\[1mm]%
\text{(ii'}_{2}\text{)} &  & \sigma^{\mathcal{O}}p_{nm}^{\mathcal{O}}%
(\theta_{1}+2\pi i,\theta_{2},\dots,\underline{z})=p_{nm}^{\mathcal{O}}%
(\theta_{1},\theta_{2},\dots,\underline{z})\\[1mm]%
\text{(iii'}_{2}\text{)} &  & \text{if }\theta_{12}=i\pi:p_{nm}^{\mathcal{O}%
}(\underline{\theta},\underline{z})|_{z_{1}=\theta_{1}}=\sigma^{\mathcal{O}%
}p_{nm}^{\mathcal{O}}(\underline{\theta},\underline{z})|_{z_{1}=\theta_{2}%
}\\[1mm]
&  & ~~~~~~~~~~~~~~~~~~~=\sigma^{\mathcal{O}}p_{n-2m-1}^{\mathcal{O}}%
(\theta_{3},\ldots,\theta_{n},z_{2},\ldots,z_{m})+\tilde{p}%
\end{array}
\]
where $\sigma^{\mathcal{O}}$ is the statistics factor of the operator
${\mathcal{O}}$ with respect to the particle. The function $\tilde{p}$ must
not contribute after integration, which means in particular that is does not
depend on the $z_{i}$ (in most cases $\tilde{p}=0$). For convenience we have
introduced the indices $_{nm}$ to denote the number of variables. For the
recursion relation (iii) in addition the normalization coefficients have to
satisfy%
\begin{equation}
N_{n}=iN_{n-2}\,.\label{n2}%
\end{equation}
One may convince oneself that the form factor satisfies (i) and\ (ii). Not so
trivial is the residue relation (iii), however, it follows from a simplified
version of the proofs for the $Z(3)$ and the general $Z(N)$ case below.

\subsection{Examples of fields and their p-functions:}

We present the following correspondence of operators and p-functions which are
solutions of (i'$_{2}$) - (iii'$_{2}$). For the order parameter $\sigma(x)$,
the disorder parameter $\mu(x)$, fermi field $\psi(x)$, and the higher
conserved currents $J_{L}^{\mu}(x),\,(L\in\mathbb{Z})$ we propose the
following p-functions and statistics parameters%
\begin{equation}%
\begin{array}
[c]{rlll}%
\sigma(x)\leftrightarrow & p_{nm}^{\sigma}(\underline{\theta},\underline
{z})=1 & ~\text{for }n=2m+1 & ~\text{with }\sigma^{\sigma}=1\\[1mm]%
\mu(x)\leftrightarrow & p_{nm}^{\mu}(\underline{\theta},\underline{z}%
)=i^{m}e^{\left(  \sum z_{i}-\frac{1}{2}\sum\theta_{i}\right)  } & ~\text{for
}n=2m & ~\text{with }\sigma^{\mu}=-1\\[1mm]%
\psi^{\pm}(x)\leftrightarrow & p_{nm}^{\psi^{\pm}}(\underline{\theta
},\underline{z})=e^{\pm\left(  \sum z_{i}-\frac{1}{2}\sum\theta_{i}\right)  }
& ~\text{for }n=2m+1 & ~\text{with }\sigma^{\psi}=-1\\[1mm]%
J_{L}^{\pm}(x)\leftrightarrow & p_{nm}^{J_{L}^{\pm}}(\underline{\theta
},\underline{z})=\sum e^{\pm\theta_{i}}\sum e^{Lz_{i}} & ~\text{for }n=2m &
~\text{with }\sigma^{\mu}=1
\end{array}
\label{2.3}%
\end{equation}
Note that $\tilde{p}\neq0$ in (iii'$_{2})$ only for $J_{L}^{\pm}$. The
motivation of these choices will be more obvious when we investigate the
commutation rules of fields in section \ref{s5} and from the properties of the
form factors which we now discuss in more detail.

\textbf{Explicit expressions of the form factors:}

Now we have to check that the proposed p-functions really provide the well
known form factors for the order, disorder and fermi fields. In order to get
simple expressions for these form factors, we have to calculate the integral
(\ref{Inm2}) with (\ref{h2}) for each p-function separately.

\textbf{For the order parameter:} Only for odd numbers of particles the form
factors are non-zero. We calculate
\begin{equation}
I_{nm}(\underline{\theta},1)=2^{m}\prod_{1\leq i<j\leq n}\frac{1}{\cosh
\frac{1}{2}\theta_{ij}}~~\,\text{for }n=2m+1. \label{2.4}%
\end{equation}
The proof of this formula can be found in Appendix \ref{ab}. The general
formulae (\ref{4}), (\ref{K}), (\ref{F2}), and (\ref{n2}) with $N_{n}%
=i^{(n-1)/2}$ then imply for $n$ odd%
\begin{equation}
F_{n}^{\sigma}(\underline{\theta})=2^{m}\prod_{1\leq i<j\leq n}\frac
{F(\theta_{ij})}{\cosh\frac{1}{2}\theta_{ij}}=\left(  2i\right)
^{(n-1)/2}\prod_{1\leq i<j\leq n}\tanh\tfrac{1}{2}\theta_{ij} \label{2.6}%
\end{equation}
which agrees with the known result (\ref{bkw}). The proof that the integral
$I_{nm}(\underline{\theta},1)$ vanishes for even $n$ and $m>0$ is simple: If
$m(m-1)<n$ we may decompose for real $\theta_{i}$ the contours $\mathcal{C}%
_{\underline{\theta}}=\mathcal{C}_{0}-\mathcal{C}_{0-i\pi}$ where
$\operatorname{Re}\mathcal{C}_{0}$ goes from $-\infty$ to $\infty$ and
$\operatorname{Im}\left(  \theta_{i}+i\pi\right)  <\operatorname{Im}%
\mathcal{C}_{0}<\operatorname{Im}\left(  \theta_{i}\right)  $. The shift
$z_{i}\rightarrow z_{i}-i\pi$ implies $h(\underline{\theta},\underline
{z})\rightarrow(-1)^{n}h(\underline{\theta},\underline{z})$ such that the
integrals along $\mathcal{C}_{0}$ and $\mathcal{C}_{0-i\pi}$ cancel for even
$n$.

\textbf{For the disorder parameter:} Only for even numbers of particles the
form factors are non-zero. We calculate with $p^{\mu}$ as given in
(\ref{2.3})
\[
I_{nm}(\underline{\theta},p^{\mu})=2^{m}\prod_{1\leq i<j\leq n}\frac{1}%
{\cosh\frac{1}{2}\theta_{ij}}~~\text{for }n=2m\,.
\]
The proof of this formula is analog to that in Appendix \ref{ab}, therefore
with $N_{n}=i^{n/2}$ the form factors are for $n$ even%
\begin{equation}
F_{n}^{\mu}(\underline{\theta})=\left(  2i\right)  ^{n/2}\prod_{1\leq i<j\leq
n}\tanh\tfrac{1}{2}\theta_{ij}\,. \label{2.7}%
\end{equation}
Similar as above for the order parameter one can show that the integral
$I_{nm}(\underline{\theta},p^{\mu})$ vanishes for odd $n$ and $m>0$. It is
also interesting to investigate the asymptotic behavior of the form factors
when one of the rapidities goes to infinity. From the integral representation
it is easy to check that
\[
F_{n}^{\sigma}(\underline{\theta})\overset{\theta_{1}\rightarrow\infty
}{\rightarrow}F_{n-1}^{\mu}(\underline{\theta}^{\prime})\overset{\theta
_{2}\rightarrow\infty}{\rightarrow}2iF_{n-2}^{\sigma}(\underline{\theta
}^{\prime\prime})\,.
\]
Of course, this result follows easily from the explicit expressions
(\ref{2.6}) and (\ref{2.7}). This asymptotic behavior is another motivation
for our choice (\ref{2.3}) of the p-function for $\sigma(x)$ and $\mu(x)$.

\textbf{For the fermi field:} Only for $n=1$ the form factors are non-zero. We
calculate with $p^{\psi^{\pm}}$ as given in (\ref{2.3})%
\[
I_{nm}(\underline{\theta},p^{\psi^{\pm}})=\delta_{n1}\,e^{\mp\frac{1}{2}%
\theta}~~\text{for }n=2m+1\,.
\]
The proof that $I_{nm}(\underline{\theta},p^{\psi^{\pm}})=0$ for $n=2m+1$ odd
and $m>0$ is the same as for the disorder parameter. Therefore with the
normalization $N_{1}=\sqrt{M}$ we obtain%
\begin{equation}
F_{1}^{\psi}(\theta)=\langle\,0\mid\psi(0)\mid\theta\,\rangle=u(\theta
)=\sqrt{M}\binom{e^{-\frac{1}{2}\theta}}{e^{\frac{1}{2}\theta}}\,.
\end{equation}
The property that all form factors of the fermi field vanish except the vacuum
one-particle matrix element reflects the fact that $\psi(x)$ is a free field,
in particular for the Wightman functions one easily obtains%
\[
\left\langle \,0\mid\psi(x_{1})\cdots\psi(x_{n})\mid0\,\right\rangle
=\left\langle \,0\mid\psi(x_{1})\cdots\psi(x_{n})\mid0\,\right\rangle
_{free}\,.
\]

\textbf{For the infinite set of conserved higher currents:} Only for $n=2$ the
form factors are non-zero. We calculate with $p^{J_{L}^{\pm}}$ as given in
(\ref{2.3})%
\[
I_{nm}(\underline{\theta},p^{J_{L}^{\pm}})=\delta_{n2}\,\left(  e^{\pm
\theta_{1}}+e^{\pm\theta_{2}}\right)  2i\left(  \frac{e^{L\theta_{1}}}%
{\sinh\theta_{12}}+\frac{e^{L\theta_{2}}}{\sinh\theta_{21}}\right)
~~\text{for }n=2m
\]
The proof that $I_{nm}(\underline{\theta},p^{J_{L}^{\pm}})=0$ for $n=2m>2$ is
again similar as above. With the normalization $c_{21}=\pm iM$ the form
factors are
\[
\langle\,0\,|\,J_{L}^{\pm}(0)\,|\,\theta_{1},\dots,\theta_{n}\rangle^{in}%
=\mp\delta_{n2}2M\left(  e^{\pm\theta_{1}}+e^{\pm\theta_{2}}\right)  \left(
e^{L\theta_{1}}-e^{L\theta_{2}}\right)  \frac{1}{\sinh\theta_{12}}%
\]
such that as in \cite{BK2} the charges $Q_{L}=\int dxJ_{L}^{0}(x)$ satisfy the
eigenvalue equation
\[
\left(  Q_{L}-\sum_{i=1}^{n}e^{L\theta_{i}}\right)  \,|\,\theta_{1}%
,\ldots,\theta_{n}\,\rangle^{in}=0~~\text{for }L=\pm1,\pm3,\dots.
\]
Obviously we get the energy momentum tensor from $J_{\pm1}^{\pm}(x)$.

The property $F_{n}^{\psi}=0$ for odd $n>1$ and $F_{n}^{J}=0$ for even $n>2$
is related to the fact that in the recursion relation (iii) the factor
$(1-\sigma^{\mathcal{O}}\prod S)$ is zero in both cases.

\section{Z(3)-form factors}

Let us now consider $N=3$, which corresponds to the scaling $Z(3)$-Ising
model. In this case we have two particles, 1 and 2, and the bound state of two
particles of type $1$ is the particle $2$, which in turn is the antiparticle
of particle $1$. Conversely, the bound state of two particles of type $2$ is
the particle of type $1$, which in turn is again the antiparticle of $2$. So,
our construction should take into account this structure of the bound states.
We construct the form factors for particles of type $1$, the others can then
be obtained by the form factor bound state formula (iv).

\subsection{The general formula for n-particle form factors}

In order to obtain a recursion relation where only form factors for particles
of type $1$ are involved, we have to apply the bound state relation (iv) to
get the anti-particle and then the creation annihilation equation (iii) to
obtain%
\begin{align}
\operatorname*{Res}_{\theta_{23}=i\eta}\operatorname*{Res}_{\theta_{12}=i\eta
}F_{1111\dots1}^{{\mathcal{O}}}(\theta_{1},\dots) &  =\operatorname*{Res}%
_{\theta_{(12)3}=i\pi}F_{211\dots1}^{{\mathcal{O}}}(\theta_{(12)},\theta
_{3},\dots)\sqrt{2}\Gamma\nonumber\\
&  =2iF_{1\dots1}^{{\mathcal{O}}}(\theta_{4},\dots)\left(  1-\sigma
_{1}^{{\mathcal{O}}}\prod_{i=4}^{n}S(\theta_{3i})\right)  \sqrt{2}%
\Gamma\label{3.2}%
\end{align}
where $\theta_{(12)}=\frac{1}{2}(\theta_{1}+\theta_{2})$ is the bound state
rapidity, $\eta=\frac{2}{3}\pi$ is the bound state fusion angle and
$\Gamma=i\left\vert \operatorname*{Res}_{\theta=i\eta}S_{11}(\theta
)\right\vert ^{1/2}$ is the bound state intertwiner (see \cite{K1,BK}). In the
following we use again the short notation $F_{1\dots1}^{{\mathcal{O}}}%
(\theta_{1},\dots,\theta_{n})=F_{n}^{{\mathcal{O}}}(\underline{\theta})$ and
also write the statistics factor as $\sigma^{{\mathcal{O}}}(1)=\sigma
_{1}^{{\mathcal{O}}}$. We write the form factors again in the form (\ref{4})
where minimal form factor function
\begin{equation}
F(\theta)=c_{3}\exp\int_{0}^{\infty}\frac{dt}{t}\,\frac{2\cosh\frac{1}%
{3}t\sinh\frac{2}{3}t}{\sinh^{2}t}\left(  1-\cosh t\left(  1-\frac{\theta
}{i\pi}\right)  \right)  \label{F3}%
\end{equation}
is the solution of Watson's equations (\ref{3}) with the S-matrix (\ref{1})
for $N=3$. The constant $c_{3}$ is given by (\ref{c}) in appendix \ref{a1}.
Similar as above we make the Ansatz for the K-functions%
\begin{equation}
K_{n}^{{\mathcal{O}}}(\underline{\theta})=N_{n}I_{nmk}(\underline{\theta
},p^{{\mathcal{O}}})\label{K3}%
\end{equation}
with the fundamental building blocks of form factors
\begin{align}
I_{nmk}(\underline{\theta},p) &  =\frac{1}{m!k!}\int_{\mathcal{C}%
_{\underline{\theta}}}\frac{dz_{1}}{R}\cdots\int_{\mathcal{C}_{\underline
{\theta}}}\frac{dz_{m}}{R}\,\int_{\mathcal{C}_{\underline{\theta}}}%
\frac{du_{1}}{R}\cdots\int_{\mathcal{C}_{\underline{\theta}}}\frac{du_{k}}%
{R}\,h(\underline{\theta},\underline{z},\underline{u})\,p(\underline{\theta
},\underline{z},\underline{u})\label{Inm3}\\
h(\underline{\theta},\underline{z},\underline{u}) &  =\prod_{i=1}^{n}\left(
\prod_{j=1}^{m}\phi(z_{j}-\theta_{i})\prod_{j=1}^{k}\phi(u_{j}-\theta
_{i})\right)  \label{h3}\\
&  \times\prod_{1\leq i<j\leq m}\tau(z_{ij})\prod_{1\leq i<j\leq k}\tau
(u_{ij})\prod_{1\leq i\leq m}\prod_{1\leq j\leq k}\varkappa(z_{i}%
-u_{j})\,.\nonumber
\end{align}
Again the integration contours $\mathcal{C}_{\underline{\theta}}%
=\sum\mathcal{C}_{\theta_{i}}$ enclose the points $\theta_{i}\,$ such that
$R=\int_{\mathcal{C}_{\theta}}dz\phi(z-\theta)$. Equations (iii) and (iv), in
particular (\ref{3.2}) lead to the relations%
\begin{gather}
\prod_{k=0}^{1}\phi(\theta+ki\eta)\prod_{k=0}^{2}F(\theta+ki\eta
)=1\label{phiF3}\\
\tau(z)\phi\left(  z\right)  \phi\left(  -z\right)  =1\ ,~~\varkappa
(z)\phi(z)=1\label{tk3}%
\end{gather}
as an extension of (\ref{phi2}) and (\ref{tau}) for the $Z(2)$ case. The
solution for $\phi$ is
\begin{equation}
\phi(z)=\frac{1}{\sinh\frac{1}{2}z\sinh\frac{1}{2}(z+i\eta)}\label{phi3}%
\end{equation}
if the constant $c_{3}$ is fixed as in (\ref{c}). The phi-function satisfies
the `Jost function' property%
\begin{equation}
\frac{\phi(-\theta)}{\phi(\theta)}=S(\theta)\,.\label{J}%
\end{equation}
We will now show again that by the Ansatz (\ref{K3})-(\ref{h3})  we have
transformed the form factor equations (i)-(v) in particular equation
(\ref{3.2}) to simple equations for the p-functions.

\noindent\textbf{Assumptions for the p-functions:} The function $p_{nmk}%
^{\mathcal{O}}(\underline{\theta},\underline{z},\underline{u})~$is analytic in
all variables and satisfies:%
\begin{equation}%
\begin{array}
[c]{lll}%
\text{(i'}_{3}\text{)} &  & p_{nmk}^{\mathcal{O}}(\underline{\theta
},\underline{z},\underline{u})~\text{is symmetric under }\theta_{i}%
\leftrightarrow\theta_{j}\\[2mm]%
\text{(ii'}_{3}\text{)} &  & \sigma_{1}^{\mathcal{O}}p_{nmk}^{\mathcal{O}%
}(\theta_{1}+2\pi i,\theta_{2},\dots,\underline{z},\underline{u}%
)=p_{nmk}^{\mathcal{O}}(\theta_{1},\theta_{2},\dots,\underline{z}%
,\underline{u})\\[2mm]%
\text{(iii'}_{3}\text{)} &  & \text{if }\theta_{12}=\theta_{23}=i\eta\\[1mm]
&  &
\begin{array}
[c]{ll}%
p_{nmk}^{\mathcal{O}}(\underline{\theta},\underline{z},\underline{u})|
_{\substack{z_{1}=\theta_{1} \\u_{1}=\theta_{2}}} & =\sigma_{1}^{\mathcal{O}%
}p_{nmk}^{\mathcal{O}}(\underline{\theta},\underline{z},\underline{u})|
_{\substack{z_{1}=\theta_{2} \\u_{1}=\theta_{3}}}\\[1mm]
& =\sigma_{1}^{\mathcal{O}}p_{n-3m-1k-1}^{\mathcal{O}}(\underline{\theta
}^{\prime},\underline{z}^{\prime},\underline{u}^{\prime})+\tilde{p}%
\end{array}
\\[5mm]%
\text{(v'}_{3}\text{)} &  & p_{nmk}^{\mathcal{O}}(\underline{\theta}%
+\mu,\underline{z}+\mu,\underline{u}+\mu)=e^{s\mu}p_{nmk}^{\mathcal{O}%
}(\underline{\theta},\underline{z},\underline{u})
\end{array}
\label{p3}%
\end{equation}
where $\underline{\theta}^{\prime}=(\theta_{4},\dots,\theta_{n}),\,\underline
{z}^{\prime}=(z_{2},\dots,z_{m})$ and $\underline{u}^{\prime}=(u_{2}%
,\dots,u_{k})$. In (ii'$_{3}$) and (iii'$_{3}$) $\sigma_{1}^{\mathcal{O}}$ is
the statistics factor of the operator ${\mathcal{O}}$ with respect to the
particle of type $1$ and in (v'$_{3}$) $s$ is the spin of the operator
${\mathcal{O}}$. Again $\tilde{p}$ must not contribute after integration (in
most cases $\tilde{p}=0$). Again one may convince oneself that the form factor
satisfies (i) and\ (ii) if $h(\underline{\theta},\underline{z})$ is symmetric
under $\theta_{i}\leftrightarrow\theta_{j}$ and periodic with respect to
$\theta_{i}\rightarrow\theta_{i}+2\pi i$. Not so trivial is the residue
relation (iii) which is proved in the following lemma.

\begin{lemma}
\label{l3}The form factors $F_{n}^{\mathcal{O}}(\underline{\theta})$ defined
by (\ref{4}) and (\ref{F3})-(\ref{h3}) satisfies \mbox{\rm(i)--(v)}, in
particular (\ref{3.2}), if the p-functions satisfy
\mbox{\rm(i'$_3$)--(v'$_5$)} of (\ref{p3}), the functions $\phi,\,\tau$ and
$\varkappa$ the relations (\ref{phiF3}), (\ref{tk3}) and the normalization
constants in (\ref{K3}) the recursion relation%
\begin{equation}
N_{n}\left(  \operatorname*{Res}_{\theta=i\eta}\phi(-\theta)\right)  ^{2}%
\phi(i\eta)\ F^{2}(i\eta)F(2i\eta)=N_{n-3}2i\sqrt{2}\Gamma\,. \label{norm3}%
\end{equation}

\end{lemma}

\begin{proof}
The form factor equations (i), (ii), and (v) follow obviously from the
equations for the p-functions (i'$_{3}$), (ii'$_{3}$), and (v'$_{3}$),
respectively. As already stated, we will prove properties (iii) and (iv)
together in the form of (\ref{3.2}). Taking the residues $\operatorname*{Res}%
_{\theta_{23}=i\eta}\operatorname*{Res}_{\theta_{12}=i\eta}$ there will be
$mk$ equal terms originating from pinchings for $z_{i}$ and $u_{i}$. We pick
them from $z_{1}$ and $u_{1}$ and rewrite the products that appear in the
expression for $I_{nmk}$ in a convenient form, such that the location of the
poles turns out to be separated from the general expression. Then, the
essential calculation to be performed is%
\begin{align*}
\operatorname*{Res}_{\theta_{23}=i\eta}\operatorname*{Res}_{\theta_{12}=i\eta
}I_{nmk}(\underline{\theta},p^{{\mathcal{O}}})  &  =\frac{mk}{m!k!}%
\int_{\mathcal{C}_{\underline{\theta}^{\prime}}}\frac{dz_{2}}{R}\cdots
\int_{\mathcal{C}_{\underline{\theta}^{\prime}}}\frac{dz_{m}}{R}%
\,\int_{\mathcal{C}_{\underline{\theta}^{\prime}}}\frac{du_{2}}{R}\cdots
\int_{\mathcal{C}_{\underline{\theta}^{\prime}}}\frac{du_{k}}{R}\\
&  \times\,\prod_{i=4}^{n}\left(  \prod_{j=2}^{m}\phi(z_{j}-\theta_{i}%
)\prod_{j=2}^{k}\phi(u_{j}-\theta_{i})\right) \\
&  \times\prod_{2\leq i<j\leq m}\tau(z_{ij})\prod_{2\leq i<j\leq k}\tau
(u_{ij})\prod_{i=2}^{m}\prod_{j=2}^{k}\varkappa(z_{i}-u_{j})\\
&  \times\prod_{i=1}^{3}\left(  \prod_{j=2}^{m}\phi(z_{j}-\theta_{i}%
)\prod_{j=2}^{k}\phi(u_{j}-\theta_{i})\right)  r
\end{align*}
with
\begin{align*}
r  &  =\operatorname*{Res}_{\theta_{23}=i\eta}\operatorname*{Res}_{\theta
_{12}=i\eta}\int_{\mathcal{C}_{\underline{\theta}}}\frac{dz_{1}}{R}%
\int_{\mathcal{C}_{\underline{\theta}}}\frac{du_{1}}{R}\prod_{i=1}^{n}\left(
\phi(z_{1}-\theta_{i})\phi(u_{1}-\theta_{i})\right) \\
&  \times\prod_{j=2}^{m}\tau(z_{1j})\prod_{j=2}^{k}\tau(u_{1j})\,\varkappa
(z_{1}-u_{1})\prod_{j=2}^{k}\varkappa(z_{1}-u_{j})\prod_{j=2}^{m}%
\varkappa(z_{j}-u_{1})\mathcal{\,}p_{n}^{{\mathcal{O}}}(\underline{\theta
},\underline{z},\underline{u}).
\end{align*}
Replacing $\mathcal{C}_{\underline{\theta}}$ by $\mathcal{C}_{\underline
{\theta}^{\prime}}$ where $\underline{\theta}^{\prime}=(\theta_{4}%
,\dots,\theta_{n})$ we have used $\tau(0)=\tau(\pm i\eta)=\varkappa
(0)=\varkappa(-i\eta)=0$ and the fact that the $z_{1},u_{1}$-integrations give
non-vanishing results only for $\left(  z_{1},u_{1}\right)  $ at $\left(
\theta_{1},\theta_{2}\right)  $ and $\left(  \theta_{2},\theta_{3}\right)  $.
This is because for $\theta_{12},\theta_{23}\rightarrow i\eta$ pinching
appears at $z_{1}=\theta_{2},\;u_{1}=\theta_{3}$ and $z_{1}=\theta_{1,\;}%
u_{1}=\theta_{2}$. Defining the function
\[
f(z_{1},u_{1})=\prod_{j=2}^{m}\tau(z_{1j})\prod_{j=2}^{k}\tau(u_{1j}%
)\varkappa(z_{1}-u_{1})\prod_{j=2}^{k}\varkappa(z_{1}-u_{j})\prod_{j=2}%
^{m}\varkappa(z_{j}-u_{1})p^{\mathcal{O}}_{n}(\underline{\theta},\underline
{z},\underline{u})
\]
and using the property that $f(z,z)=f(z,z-i\eta)=0$, we calculate
\begin{multline*}
r=\operatorname*{Res}_{\theta_{23}=i\eta}\operatorname*{Res}_{\theta
_{12}=i\eta}\int_{\mathcal{C}_{\underline{\theta}}}\frac{dz_{1}}{R}%
\int_{\mathcal{C}_{\underline{\theta}}}\frac{du_{1}}{R}f(z_{1},u_{1}%
)\prod_{i=1}^{n}\left(  \phi(z_{1}-\theta_{i})\phi(u_{1}-\theta_{i})\right) \\
=\left(  \operatorname*{Res}_{\theta=i\eta}\phi(-\theta)\right)  ^{2}\phi
^{2}(i\eta)\prod_{i=4}^{n}\left(  \phi(\theta_{2i})\phi(\theta_{3i})\right)
\left(  f(\theta_{2},\theta_{3})-f(\theta_{1},\theta_{2})\prod_{i=4}%
^{n}S(\theta_{3i})\right)
\end{multline*}
We have used the symmetries of the phi-function $\phi(\theta)=\phi(\theta+2\pi
i)=\phi(-\theta-i\eta)$ the Jost property (\ref{J}) and $\theta_{12}%
,\theta_{23}=i\eta=\frac{2}{3}i\pi$ which implies that%
\begin{align*}
\operatorname*{Res}_{\theta_{12}=i\eta}\phi(\theta_{21})\operatorname*{Res}%
_{\theta_{23}=i\eta}\phi(\theta_{32})  &  =\operatorname*{Res}_{\theta
_{12}=i\eta}\phi(\theta_{13})\operatorname*{Res}_{\theta_{23}=i\eta}%
\phi(\theta_{21})=\left(  \operatorname*{Res}_{\theta=i\eta}\phi
(-\theta)\right)  ^{2}\\
\phi(\theta_{23})\phi(\theta_{31})  &  =\phi(\theta_{12})\phi(\theta
_{23})=\phi^{2}(i\eta)\\
\frac{\phi(\theta_{1i})}{\phi(\theta_{3i})}  &  =\frac{\phi(\theta_{3i}%
-i\eta)}{\phi(\theta_{3i})}=\frac{\phi(-\theta_{3i})}{\phi(\theta_{3i}%
)}=S(\theta_{3i})\,.
\end{align*}
With the help of the defining equations (\ref{tk3}) for $\tau$ and $\varkappa$
which imply%
\begin{align*}
\left(  \prod_{i=1}^{3}\phi(z-\theta_{i})\right)  ^{-1}  &  =\tau(\theta
_{1}-z)\varkappa(z-\theta_{2})=\tau(\theta_{2}-z)\varkappa(\theta_{1}-z)\\
&  =\tau(\theta_{2}-z)\varkappa(z-\theta_{3})=\tau(\theta_{3}-z)\varkappa
(\theta_{1}-z)
\end{align*}
we obtain the relations for $f(\theta_{2},\theta_{3})$ and $f(\theta
_{1},\theta_{2})$%
\begin{align*}
\prod_{i=1}^{3}\left(  \prod_{j=2}^{m}\phi(z_{j}-\theta_{i})\prod_{j=2}%
^{k}\phi(u_{j}-\theta_{i})\right)  f(\theta_{1},\theta_{2})  &  =\varkappa
(\theta_{12})p_{n}^{{\mathcal{O}}}(\underline{\theta},\theta_{1},\underline
{z}^{\prime},\theta_{2},\underline{u}^{\prime})\\
\prod_{i=1}^{3}\left(  \prod_{j=2}^{m}\phi(z_{j}-\theta_{i})\prod_{j=2}%
^{k}\phi(u_{j}-\theta_{i})\right)  f(\theta_{2},\theta_{3})  &  =\varkappa
(\theta_{23})p_{n}^{{\mathcal{O}}}(\underline{\theta},\theta_{2},\underline
{z}^{\prime},\theta_{3},\underline{u}^{\prime})\,.
\end{align*}
Finally we obtain using the defining relation (\ref{phiF3}) for the
phi-function%
\begin{align*}
\operatorname*{Res}_{\theta_{23}=i\eta}\operatorname*{Res}_{\theta_{12}=i\eta
}I_{nmk}(\underline{\theta},p^{\mathcal{O}})  &  =\left(  \operatorname*{Res}%
_{\theta=i\eta}\phi(-\theta)\right)  ^{2}\phi^{2}(i\eta)\varkappa(i\eta
)\prod_{i=1}^{3}\prod_{j=4}^{n}\left(  F(\theta_{ij})\right)  ^{-1}\\
&  \times I_{n-3m-1k-1}(\underline{\theta}^{\prime},p^{\mathcal{O}})\left(
1-\sigma_{1}^{\mathcal{O}}\prod_{i=4}^{n}S(\theta_{3i})\right)
\end{align*}
if the p-function satisfies (iii'$_{3}$). Therefore the form factor given by
(\ref{4}) and (\ref{F3})--(\ref{h3}) satisfies (\ref{2.3}) and (iii) if the
normalization constants satisfy (\ref{norm3}).
\end{proof}

\subsection{Examples of fields and p-functions}

We present solutions of the equations for the p-functions (i'$_{3}$%
)--(v'$_{3}$) of (\ref{p3}) and some explicit examples of the resulting form
factors. We identify the fields by the properties of their matrix elements. In
section \ref{s5} we show that the field satisfy the desired commutation rules.
This motivates to propose a correspondence of fields $\phi(x)$ and p-functions
$p_{nmk}^{\phi}(\underline{\theta},\underline{z},\underline{u})$.

\noindent\textbf{The order parameter $\sigma_{Q}(x)$:} We look for a solution
of (i'$_{3}$)-(v'$_{3}$) with
\[
\left\{
\begin{array}
[c]{lrcl}%
\text{charge} & Q & = & 1,2\\
\text{spin} & s & = & 0\\
\text{statistics} & \sigma_{1}^{\sigma_{Q}} & = & 1\,.
\end{array}
\right.
\]
Since the fields carry the charge $Q$ the only non-vanishing form factors with
$n$ particles of type 1 are the ones with $n=Q\operatorname{mod}3$. We propose
the correspondence of the field and the p-function:%
\[
\sigma_{Q}\leftrightarrow p_{nmk}^{\sigma_{Q}}=1\quad\text{with }%
n=3l+Q,\,\left\{
\begin{array}
[c]{llc}%
m=l+1 & k=l & \text{for }Q=1\\
m=l+1 & k=l+1 & \text{for }Q=2\,.
\end{array}
\right.
\]
The normalization constants $N_{n}$ follow from (\ref{norm3}).

\textbf{Examples for $Q=1$:} The form factors of the order parameter
$\sigma_{1}(x)$ for one and four particles of type 1 are%
\begin{align*}
F_{1}^{\sigma_{1}}  &  =\langle\,0\,|\,\sigma_{1}(0)\,|\,p\rangle_{1}%
=N_{1}I_{110}=1\\
F_{1111}^{\sigma_{1}}(\underline{\theta})  &  =\langle\,0\,|\,\sigma
_{1}(0)\,|\,p_{1},p_{2},p_{3},p_{4}\rangle_{1111}^{in}=N_{4}I_{421}%
(\underline{\theta},1)\prod_{1\leq i<j\leq4}F(\theta_{ij})
\end{align*}
where we calculate from our integral representation (\ref{Inm3})%
\[
I_{421}(\underline{\theta},1)=const\left(  \sum e^{-\theta_{i}}\sum
e^{\theta_{i}}-1\right)  \prod\limits_{i<j}\frac{1}{\sinh\frac{1}{2}%
(\theta_{ij}-i\eta)\sinh\frac{1}{2}\left(  \theta_{ij}+i\eta\right)  }\,.
\]
This result has already been obtained in \cite{K3} where also the form factor
equation (iv) has been discussed, in particular (up to normalizations)%
\[
\operatorname*{Res}\limits_{\theta_{34}=2\pi i/3}\langle\,0\,|\,\sigma
_{1}(0)\,|\,p_{1},p_{2},p_{3},p_{4}\,\rangle_{1111}^{in}=const\langle
\,0\,|\,\sigma_{1}(0)\,|\,p_{1},p_{2},p_{3}+p_{4}\,\rangle_{112}^{in}%
\]
with%
\[
\langle\,0\,|\,\sigma_{1}(0)\,|\,p_{1},p_{2},p_{3}\,\rangle_{112}^{in}%
=\frac{const\ F(\theta_{12})}{\sinh\frac{1}{2}(\theta_{12}-i\eta)\sinh\frac
{1}{2}\left(  \theta_{12}+i\eta\right)  }\prod_{i=1}^{2}\frac{F_{12}^{\min
}(\theta_{i3})}{\cosh\frac{1}{2}\theta_{i3}}%
\]
where $F_{12}^{\min}$ is the minimal form factor function for the S-matrix
$S_{12}$. Further it has been found in \cite{K3} that
\[
\operatorname*{Res}\limits_{\theta_{12}=2\pi i/3}\langle\,0\,|\,\sigma
_{1}(0)\,|\,p_{1},p_{2},p_{3}\,\rangle_{112}^{in}=const\langle\,0\,|\,\sigma
_{1}(0)\,|\,p_{1}+p_{2},p_{3}\,\rangle_{22}^{in}%
\]
with%
\begin{equation}
\langle\,0\,|\,\sigma_{1}(0)\,|\,p_{1},p_{2}\,\rangle_{22}^{in}=\frac
{const\ F(\theta_{12})}{\sinh\frac{1}{2}(\theta_{12}-i\eta)\sinh\frac{1}%
{2}\left(  \theta_{12}+i\eta\right)  } \label{3.3}%
\end{equation}
and the form factor equation (iii) has been checked%
\[
\operatorname*{Res}\limits_{\theta_{23}=i\pi}\langle\,0\,|\,\sigma
_{1}(0)\,|\,p_{1},p_{2},p_{3}\,\rangle_{112}^{in}=const\,\left(  S(\theta
_{12})-1\right)  \,.
\]

\textbf{Example for $Q=2$:} The form factor of the order parameter $\sigma
_{2}(x)$ for two particles of type 1 is
\[
F_{11}^{\sigma_{2}}(\underline{\theta})=\langle\,0\,|\,\sigma_{2}%
(0)\,|\,p_{1},p_{2}\rangle_{11}^{in}=N_{2}I_{211}(\underline{\theta
},1)F(\theta_{12})
\]
where we calculate
\[
I_{211}(\underline{\theta},1)=const\frac{1}{\sinh\frac{1}{2}(\theta_{12}%
-i\eta)\sinh\frac{1}{2}(\theta_{12}+i\eta)}%
\]
which agrees with the result (\ref{3.3}) of \cite{K3}. This is to be expected
because of charge conjugation.

\noindent\textbf{The disorder parameter $\mu_{\tilde{Q}}(x)$:} We look for a
solution of (i'$_{3}$)-(v'$_{3}$) with
\[
\left\{
\begin{array}
[c]{lrcl}%
\text{charge} & Q & = & 0\\
\text{spin} & s & = & 0\\
\text{statistics} & \sigma_{_{1}}^{\mu_{\tilde{Q}}} & = & \omega^{\tilde{Q}}%
\end{array}
\right.
\]
where $\omega=e^{i\eta},\eta=2/3$. We call the number $\tilde{Q}=1,2$ the
`dual charge' of the field $\mu_{\tilde{Q}}(x)$. Since the fields carry the
charge $Q=0$ the only non-vanishing form factors with $n$ particles of type 1
are the ones with $n=0\operatorname{mod}3$. We propose the correspondence of
the field and the p-function:$\tilde{Q}$%
\[
\mu_{\tilde{Q}}\leftrightarrow\left\{
\begin{array}
[c]{l}%
p_{nmk}^{\mu_{1}}=\rho\exp\left(  \sum\limits_{i=1}^{m}z_{i}-\frac{1}{3}%
\sum\limits_{i=1}^{n}\theta_{i}\right) \\
p_{nmk}^{\mu_{2}}=\rho\exp\left(  \sum\limits_{i=1}^{m}z_{i}+\sum
\limits_{i=1}^{k}u_{i}-\frac{2}{3}\sum\limits_{i=1}^{n}\theta_{i}\right)
\end{array}
\right.  \quad\text{with }\left\{
\begin{array}
[c]{l}%
n=3m\\
\,k=m
\end{array}
\right.
\]
where $\rho=\sqrt{\omega}^{\tilde{Q}(\tilde{Q}-N+2)m}$. Again the
normalization constants $N_{n}$ follow from (\ref{norm3}).

\textbf{Examples for $\tilde{Q}=1,2$:} The form factors of the disorder
parameter $\mu_{\tilde{Q}}(x)$ for 0 and 3 particles of type 1 are%
\begin{align*}
F^{\mu_{\tilde{Q}}}  &  =\langle\,0\,|\,\mu_{\tilde{Q}}(0)\,|\,0\rangle=1\\
F_{111}^{\mu_{\tilde{Q}}}(\underline{\theta})  &  =\langle\,0\,|\,\mu
_{\tilde{Q}}(0)\,|\,p_{1},p_{2},p_{3}\rangle_{111}^{in}=N_{3}I_{311}%
(\underline{\theta},p^{\mu_{\tilde{Q}}})\prod_{1\leq i<j\leq3}F_{11}%
(\theta_{ij})\,.
\end{align*}
We calculate from our integral representation (\ref{Inm3})%
\[
I_{311}(\underline{\theta},p^{\mu_{\tilde{Q}}})=const\,e^{\mp\frac{1}{3}%
\sum\limits_{i=1}^{n}\theta_{i}}\sum_{i=1}^{3}e^{\pm\theta_{i}}\prod
\limits_{i<j}\frac{1}{\sinh\frac{1}{2}(\theta_{ij}-i\eta)\sinh\frac{1}%
{2}\left(  (\theta_{ij}+i\eta\right)  }%
\]
where the upper sign is for $\tilde{Q}=1$ and the lower one for $\tilde{Q}=2$.
Using the form factor bound state formula (iv) we obtain (up to a constant)%
\[
F_{12}^{\mu_{\tilde{Q}}}(\underline{\theta})=e^{\mp\frac{1}{6}\theta_{12}%
}\frac{1}{\cosh\frac{1}{2}\theta_{12}}F_{12}^{\min}(\theta_{12}).
\]
It is interesting to note that for $\operatorname{Re}\theta_{1}\rightarrow
\infty$ we have the relation of order and disorder parameter form factors (up
to constants)%
\[
\lim_{\operatorname{Re}\theta_{1}\rightarrow\infty}\langle\,0\,|\,\sigma
_{1}(0)\,|\,p_{1},p_{2},p_{3},p_{4}\rangle_{1111}^{in}=\langle\,0\,|\,\mu
_{2}(0)\,|\,p_{2},p_{3},p_{4}\rangle_{111}^{in}%
\]
which follows from the asymptotic behavior
\begin{align*}
F(\theta_{1i})  &  \rightarrow e^{\frac{2}{3}\theta_{1i}}\\
I_{421}(\underline{\theta},1)  &  \rightarrow e^{\theta_{1}}\left(  \sum
_{i=2}^{4}e^{-\theta_{i}}\right)  \prod_{j=2}^{4}e^{-\theta_{1j}}\prod
_{1<i<j}\frac{1}{\sinh\frac{1}{2}(\theta_{ij}-a)\sinh\frac{1}{2}(\theta
_{ij}+a)}.
\end{align*}

\noindent\textbf{The para-fermi field $\psi_{Q}(x)$:} We look for a solution
of (i'$_{3}$)-(v'$_{3}$) with
\[
\left\{
\begin{array}
[c]{llll}%
\text{charge} & Q & = & Q\\
\text{spin} & s & = & Q(3-Q)/3\\
\text{statistics~} & \sigma_{_{1}}^{\psi_{Q}} & = & \omega^{Q}\,.
\end{array}
\right.
\]
These fields have charge $Q=1,2$ and dual charge $\tilde{Q}=Q$. The only
non-vanishing form factors with $n$ particles of type 1 are the ones with
$n=Q\operatorname{mod}3$. We propose the correspondence of the field and the
p-function:%
\[
\psi_{Q}\leftrightarrow\left\{
\begin{array}
[c]{l}%
p_{nmk}^{\psi_{1}}=\rho\exp\left(  \sum\limits_{i=1}^{m}z_{i}-\frac{1}{3}%
\sum\limits_{i=1}^{n}\theta_{i}\right)  \\
p_{nmk}^{\psi_{2}}=\rho\exp\left(  \sum\limits_{i=1}^{m}z_{i}+\sum
\limits_{i=1}^{k}u_{i}-\frac{2}{3}\sum\limits_{i=1}^{n}\theta_{i}\right)
\end{array}
\right.  \quad\text{with }\left\{
\begin{array}
[c]{l}%
n=3l+Q\\
m=l+1\\
k=l+Q-1
\end{array}
\right.
\]
where $\rho=\sqrt{\omega}^{\tilde{Q}(\tilde{Q}-1)l}$. Again the normalization
constants $N_{n}$ follow from (\ref{norm3}).

\textbf{Examples for $Q=1,2$:} The form factors of the para-fermi field
$\psi_{1}(x)$ for 1 and 4 particles of type 1 are%
\begin{align*}
F_{1}^{\psi_{1}}(\theta)  &  =\langle\,0\,|\,\psi_{1}(0)\,|\,p\rangle
_{1}=N_{1}I_{110}(\theta,p^{\psi_{1}})=e^{\frac{2}{3}\theta}\\
F_{1111}^{\psi_{1}}(\underline{\theta})  &  =\langle\,0\,|\,\psi
_{1}(0)\,|\,p_{1},p_{2},p_{3},p_{4}\rangle_{1111}^{in}=N_{4}I_{421}%
(\underline{\theta},p^{\psi_{1}})\prod_{1\leq i<j\leq4}F(\theta_{ij})\\
&  =const\,e^{-\frac{2}{3}\sum\limits_{i=1}^{4}\theta_{i}}\sum_{i<j}%
e^{\theta_{i}+\theta_{j}}\prod_{1\leq i<j\leq4}\frac{F(\theta_{ij})}%
{\sinh\frac{1}{2}(\theta_{ij}-i\eta)\sinh\frac{1}{2}(\theta_{ij}+i\eta)}%
\end{align*}
and the one of the para-fermi field $\psi_{2}(x)$ for 2 particles of type 1 is%
\begin{align*}
F_{11}^{\psi_{2}}(\underline{\theta})  &  =\langle\,0\,|\,\psi_{2}%
(0)\,|\,p_{1},p_{2}\rangle_{11}=N_{2}I_{211}(\underline{\theta},p^{\psi_{2}%
})F(\theta_{12})\\
&  =const\,e^{\frac{1}{3}\left(  \theta_{1}+\theta_{2}\right)  }\frac
{F(\theta_{12})}{\sinh\frac{1}{2}(\theta_{12}-i\eta)\sinh\frac{1}{2}%
(\theta_{12}+i\eta)}\,.
\end{align*}
All these examples agree with the results of \cite{KiSm}.

\noindent\textbf{The higher currents $J_{L}^{\pm}(x)$:} We look for a solution
of (i'$_{3}$)-(v'$_{3}$) with
\[
\left\{
\begin{array}
[c]{llll}%
\text{charge} & Q & = & 0\\
\text{spin} & s & = & L\pm1\\
\text{statistics~} & \sigma_{_{1}}^{J_{L}^{\pm}} & = & 1\,.
\end{array}
\right.
\]
Since the currents are $Z(3)$-charge-less the only non-vanishing form factors
with $n$ particles of type 1 are the ones with $n=0\operatorname{mod}3$. We
propose the correspondence of the currents and the p-functions for
$L\in\mathbb{Z}$%
\[
J_{L}^{\pm}\leftrightarrow p_{nmk}^{J_{L}^{\pm}}=\pm\left(  \sum_{i=1}%
^{n}e^{\pm\theta_{i}}\right)  \left(  \sum_{i=1}^{m}e^{Lz_{i}}+\sum_{i=1}%
^{m}e^{Lu_{i}}\right)  \quad\text{for }\left\{
\begin{array}
[c]{l}%
n=3m\\
k=m\,.
\end{array}
\right.
\]
Note that for this case the function $\tilde{p}$ in (\ref{p3}) is
non-vanishing, however, it does not contribute because $I_{nmm}(\underline
{\theta},1)=0$ for $n=3m$. The proof of this fact is similar to that one in
appendix \ref{ab}. The higher charges $Q_{L}=\int dxJ_{L}^{0}(x)$ satisfy the
eigenvalue equations%
\[
\left(  Q_{L}-\sum_{i=1}^{n}e^{L\theta_{i}}\right)  \,|\,p_{1},\ldots
,p_{n}\,\rangle^{in}=0\,.
\]
Obviously, from $J_{\pm1}^{\pm}(x)$ we obtain the energy momentum tensor.

\textbf{Examples:} The form factors of energy momentum tensor that is
$J_{L}^{\pm}(x)$ for $L=\pm1$ for 0 and 3 particles of type 1 are
\begin{align*}
F^{J_{L}^{\pm}}  &  =\langle\,0\,|\,J_{L}^{\pm}(0)|\,0\rangle=0\\
F_{111}^{J_{L}^{\pm}}(\underline{\theta})  &  =\langle\,0\,|\,J_{L}^{\pm
}(0)\,\,|\,p_{1},p_{2},p_{3}\rangle_{111}^{in}=c_{311}I_{311}(\underline
{\theta},p^{J_{L}^{\pm}})\prod_{1\leq i<j\leq3}F(\theta_{ij})\\
&  =\pm const\,\left(  e^{\pm\theta_{1}}+e^{\pm\theta_{2}}+e^{\pm\theta_{3}%
}\right)  \left(  e^{L\theta_{1}}+e^{L\theta_{2}}+e^{L\theta_{3}}\right) \\
&  \times\prod_{1\leq i<j\leq3}\frac{F(\theta_{ij})}{\sinh\frac{1}{2}%
(\theta_{ij}-i\eta)\sinh\frac{1}{2}(\theta_{ij}+i\eta)}\,.
\end{align*}
By (iv) we obtain the bound state form factor (up to a normalization) for
$L=\pm1$
\[
F_{12}^{J_{L}^{\pm}}(\underline{\theta})=\pm const\,e^{\frac{1}{2}\left(
L\pm1\right)  \left(  \theta_{1}+\theta_{2}\right)  }F_{12}^{\min}(\theta
_{12})\,.
\]
Notice that this last expression agrees with the results of \cite{JKOPS} when
$N=3$.

\section{Z(N)-form factors\label{s4}}

The scaling $Z(N)$-Ising model possesses particles of type $\alpha
=1,\dots,N-1$ with $Z(N)$-charge $Q_{\alpha}=\alpha$ such that the anti
particle of $\alpha$ is $\bar{\alpha}=N-\alpha$. The bound state fusion rules
are $(\alpha\beta)=\alpha+\beta\operatorname{mod}N$, in particular the bound
state of $N-1$ particles of type $1$ is the antiparticle $\bar{1}$. Therefore
applying $N-2$ times formula (iv) and once (iii) we obtain the recursion
relations for form factors where only particles of type 1 are involved%
\begin{multline}
\operatorname*{Res}_{\theta_{N-1N}=i\eta}\dots\operatorname*{Res}_{\theta
_{12}=i\eta}F_{n}(\theta_{1},\dots,\theta_{n})=2iF_{n-N}(\theta_{N+1}%
,\dots,\theta_{n})\label{4.3}\\
\times\left(  1-\sigma_{1}^{{\mathcal{O}}}\prod_{i=N+1}^{n}S(\theta
_{Ni})\right)  \prod\limits_{\alpha=1}^{N-2}\sqrt{2}\Gamma_{1\alpha}%
^{1+\alpha}%
\end{multline}
where $\eta=\frac{2\pi i}{N}$ and the $\Gamma_{1\alpha}^{1+\alpha}=i\left\vert
\operatorname*{Res}_{\theta=i\eta_{\alpha}}S_{1\alpha}(\theta)\right\vert
^{1/2}$ are the bound state intertwiners of the fusion $(1\alpha)=1+\alpha$.
We will construct the form factors of particles of type 1, all the others are
then obtained by the bound state formula (iv).

\subsection{The general Z(N)-form factor formula}

Following \cite{KW} we write the form factors again in the form (\ref{4})
\begin{equation}
F_{\underline{\alpha}}^{\mathcal{O}}(\underline{\theta})=K_{\underline{\alpha
}}^{\mathcal{O}}(\underline{\theta})\prod_{1\leq i<j\leq n}F(\theta_{ij})
\label{FO}%
\end{equation}
where minimal form factor function \cite{K3}%
\begin{equation}
F(\theta)=c_{N}\exp\int_{0}^{\infty}\frac{dt}{t}\,\frac{2\cosh\frac{1}%
{N}t\sinh\frac{N-1}{N}t}{\sinh^{2}t}\left(  1-\cosh t\left(  1-\frac{\theta
}{i\pi}\right)  \right)  \label{FN}%
\end{equation}
is the solution of Watson's equations (\ref{3}) with the S-matrix (\ref{1}).
The constant $c_{N}$ is given by (\ref{c}) in appendix \ref{a1}. The
K-function $K_{n}^{{\mathcal{O}}}(\theta_{1},\dots,\theta_{n})$ is totally
symmetric in the rapidities $\theta_{i}$, $2\pi i$ periodic, containing the
entire pole structure and determines the asymptotic behavior for large values
of the rapidities. Similar as above we make the Ansatz for the K-functions%
\begin{equation}
K_{n}^{{\mathcal{O}}}(\underline{\theta})=N_{n}I_{n\underline{m}}%
(\underline{\theta},p^{{\mathcal{O}}}) \label{KN}%
\end{equation}
with the fundamental building blocks of form factors%
\begin{align}
I_{n\underline{m}}(\underline{\theta},p^{\mathcal{O}})  &  =\left(
\prod_{k=1}^{N-1}\frac{1}{m_{k}!}\prod_{j=1}^{m_{k}}\int_{\mathcal{C}%
_{\underline{\theta}}}\frac{dz_{kj}}{R}\right)  h(\underline{\theta
},\underline{z})p^{\mathcal{O}}(\underline{\theta},\underline{z})\label{Inm}\\
h(\underline{\theta},\underline{z})  &  =\prod_{k=1}^{N-1}\left(  \prod
_{j=1}^{m_{k}}\prod_{i=1}^{n}\phi(z_{kj}-\theta_{i})\prod_{1\leq i<j\leq
m_{k}}\tau(z_{ki}-z_{kj})\right) \label{h}\\
&  \times\prod_{1\leq k<l\leq N-1}\prod_{i=1}^{m_{k}}\prod_{j=1}^{m_{l}%
}\varkappa(z_{ki}-z_{lj})\nonumber
\end{align}
where $\underline{m}=(m_{1},\dots,m_{N-1})$ and $\underline{z}=(z_{ki}%
),\,k=1,\dots,N-1,\,i=1,\dots,m_{k}$. Again the integration contours
$\mathcal{C}_{\underline{\theta}}=\sum\mathcal{C}_{\theta_{i}}$ enclose the
points $\theta_{i}\,$ such that $R=\int_{\mathcal{C}_{\theta}}dz\phi
(z-\theta)$. Equations (iii) and (iv), in particular (\ref{4.3}) lead to the
relations
\begin{gather}
\prod_{k=0}^{N-2}\phi(z+ki\eta)\prod_{k=0}^{N-1}F(z+ki\eta)=1\label{phiF}\\
\tau(-z)\phi\left(  z+i\pi\right)  \phi\left(  z\right)  =1\ ,~~\varkappa
(z)\phi(z)=1\label{tk}\\
N_{n}\left(  \operatorname*{Res}_{\theta=i\eta}\phi(-\theta)\right)
^{N-1}\prod_{k=1}^{N-2}\phi^{k}(ki\eta)\ \prod_{k=1}^{N-1}F^{N-k}%
(ki\eta)=N_{n-N}2i\prod\limits_{k=1}^{N-2}\prod\limits_{\alpha=1}^{N-2}%
\sqrt{2}\Gamma_{1\alpha}^{1+\alpha} \label{n}%
\end{gather}
where $\underline{m}-1=(m_{1}-1,\dots,m_{N-1}-1)$. The solution of
(\ref{phiF}) for $\phi$ is again%
\begin{equation}
\phi(z)=\frac{1}{\sinh\frac{1}{2}z\sinh\frac{1}{2}(z+i\eta)} \label{phi}%
\end{equation}
if the constant $c_{N}$ is fixed as in (\ref{c}). The phi-function satisfies
again the `Jost function' property $\phi(-\theta)/\phi(\theta)=S(\theta)$. The
\textbf{p-function} $p_{n\underline{m}}^{\mathcal{O}}(\underline{\theta
},\underline{z})~$is analytic in all variables and satisfies:%
\begin{equation}%
\begin{array}
[c]{lll}%
\text{(i')} &  & p_{n\underline{m}}^{\mathcal{O}}(\underline{\theta
},\underline{z})~\text{is symmetric under }\theta_{i}\leftrightarrow\theta
_{j}\\[2mm]%
\text{(ii')} &  & \sigma_{1}^{\mathcal{O}}p_{n\underline{m}}^{\mathcal{O}%
}(\theta_{1}+2\pi i,\theta_{2},\dots,\underline{z})=p_{n\underline{m}%
}^{\mathcal{O}}(\theta_{1},\theta_{2},\dots,\underline{z})\\[2mm]%
\text{(iii')} &  & \text{if }\theta_{kk+1}=i\eta=2\pi i/N~~\text{for
}k=1,\dots,N-1\\[1mm]
&  & ~~%
\begin{array}
[c]{ll}%
p_{n\underline{m}}^{\mathcal{O}}(\underline{\theta},\underline{z}%
)|_{z_{k1}=\theta_{k}} & =\sigma_{1}^{\mathcal{O}}p_{n\underline{m}%
}^{\mathcal{O}}(\underline{\theta},\underline{z})|_{z_{k1}=\theta_{k+1}%
}\\[2mm]
& =\sigma_{1}^{\mathcal{O}}p_{n-N\underline{m}-1}^{\mathcal{O}}(\underline
{\theta}^{\prime},\underline{z}^{\prime})+\tilde{p}%
\end{array}
\\[6mm]%
\text{(v')} &  & p_{nmk}^{\mathcal{O}}(\underline{\theta}+\mu,\underline
{z}+\mu)=e^{s\mu}p_{nmk}^{\mathcal{O}}(\underline{\theta},\underline{z})
\end{array}
\label{p}%
\end{equation}
where $\underline{\theta}^{\prime}=(\theta_{N+1},\dots,\theta_{n}%
),\,\underline{z}^{\prime}=(z_{ki}),\,k=1,\dots,N-1,\,i=2,\dots,m_{k}$. In
(ii') and (iii') $\sigma_{1}^{\mathcal{O}}$ is the statistics factor of the
operator ${\mathcal{O}}$ with respect to the particle of type $1$ and in (v')
$s$ is the spin of the operator ${\mathcal{O}}$. Again $\tilde{p}$ must not
contribute after integration (in most cases $\tilde{p}=0$).

By means of the off-shell Bethe Ansatz (\ref{K}) and (\ref{Inm}) we have
transformed the complicated form factor equations (i) -- (v) to simple
equations for the p-functions (i') -- (v'). Again one may convince oneself
that the form factor satisfies (i) and\ (ii) if $h(\underline{\theta
},\underline{z})$ is symmetric under $\theta_{i}\leftrightarrow\theta_{j}$ and
periodic with respect to $\theta_{i}\rightarrow\theta_{i}+2\pi i$. Not so
trivial is again the residue relation (iii) which is proved in the following
lemma.

\begin{lemma}
\label{l4}The form factors given by equations (\ref{FO}) -- (\ref{h}) satisfy
the form factor equations \mbox{\rm(i) -- (v)} if the functions $\phi
,\tau,\varkappa$ satisfy (\ref{phiF}) and (\ref{tk}), the normalization
constants satisfy (\ref{n}) and the p-functions satisfy \mbox{\rm(i') --
(v')} of (\ref{p}).
\end{lemma}

The proof of this lemma follows the same strategy of the previous $Z(3)$ case.
Here, however, the essential calculation is much more involved, due to the
existence more types of particles. Details of this proof can be found in
Appendix \ref{a2}.

\subsection{Examples of fields and p-functions}

We present solutions of the equations for the p-functions (i')--(v') of
(\ref{p}) and some explicit examples of the resulting form factors. We
identify the fields by the properties of their matrix elements. In section
\ref{s5} we show that the fields satisfy the desired commutation rules. This
motivates to propose a correspondence of fields $\phi(x)$ and p-functions
$p^{\phi}(\underline{\theta},\underline{z})$.

\noindent\textbf{The fields $\psi_{Q,\tilde{Q}}(x)$:} These fields have the
charge $Q=0,\dots,N-1$ and the dual charge $\tilde{Q}=0,\dots,N-1$. We look
for a solution of (i')-(v') with%

\begin{equation}
\left\{
\begin{array}
[c]{ll}%
\text{charge} & Q\operatorname{mod}N\\
\text{spin} & s^{\psi}=\min(Q,\tilde{Q})-Q\tilde{Q}/N\\
\text{statistics} & \sigma_{1}^{\psi}=\omega^{\tilde{Q}}%
\end{array}
\right.  \label{r}%
\end{equation}
with $\omega=e^{i\eta}=e^{2\pi i/N}$. The phase factor $\sigma_{1}^{\psi}$ is
the statistics factor of the field $\psi_{Q,\tilde{Q}}(x)$ with respect to the
particle of type 1. Since the fields carry the charge $Q$ the only
non-vanishing form factors with $n$ particles of type 1 are the ones with
$n=Q\operatorname{mod}N$. We propose the correspondence of the field and the
p-function:%
\begin{equation}
\psi_{Q,\tilde{Q}}\leftrightarrow p_{n\underline{m}}^{Q\tilde{Q}}=\rho
\exp\left(  \sum_{k=1}^{\tilde{Q}}\sum_{j=1}^{m_{k}}z_{kj}-\frac{\tilde{Q}}%
{N}\sum_{i=1}^{n}\theta_{i}\right)  \label{4.2}%
\end{equation}%
\[
\text{with }n=3l+Q,~l=0,1,2,\dots~\text{and }\left\{
\begin{array}
[c]{ll}%
m_{k}=l+1 & \text{for }k\leq Q\\
m_{k}=l & \text{for }Q<k
\end{array}
\right.
\]
where $\rho=\sqrt{\omega}^{\tilde{Q}(\tilde{Q}-N+2)n/N}$. One easily checks
that this p-function satisfies the equations (i')-(v') and the requirements
(\ref{r}). The normalization constants $N_{n}$ follow from (\ref{n}). In
particular we have for
\[%
\begin{tabular}
[c]{lll}%
$\tilde{Q}=0$ & the order parameters & $\sigma_{Q}(x)=\psi_{Q0}(x)$\\
$Q=0$ & the disorder parameter & $\mu_{\tilde{Q}}(x)=\psi_{0\tilde{Q}}(x)$\\
$Q=\tilde{Q}$ & the para-fermi fields & $\psi_{Q}(x)=\psi_{QQ}(x)\,.$%
\end{tabular}
\]
They satisfy space like commutation rules (\ref{cr}), derived in the next
section. The para-fermi fields $\psi_{Q}(x)$ are the massive analogs of the
para-fermi fields in the conformal quantum field theory of \cite{Za,Fa}. One
obtains a second set of fields $\tilde{\psi}_{Q,\tilde{Q}}(x)$ by changing the
sign in the exponent of (\ref{4.2}).

\noindent\textbf{The higher currents $J_{L}^{\pm}(x)$:} These fields are
charge-less, have bosonic statistics and spin $L\pm1$. The only non-vanishing
form factors with $n$ particles of type 1 are the ones with
$n=0\operatorname{mod}N$. We propose the correspondence of the currents and
the p-functions for $L\in\mathbb{Z}$%
\[
J_{L}^{\pm}\leftrightarrow p_{n\underline{m}}^{J_{L}^{\pm}}=\pm\sum_{i=1}%
^{n}e^{\pm\theta_{i}}\sum_{k=1}^{N-1}\sum_{j=1}^{m}e^{Lz_{kj}}\quad\text{for
}n=3m\,.
\]
The higher charges $Q_{L}=\int dxJ_{L}^{0}(x)$ satisfy again the eigenvalue
equations%
\[
\left(  Q_{L}-\sum_{i=1}^{n}e^{L\theta_{i}}\right)  \,|\,p_{1},\ldots
,p_{n}\,\rangle^{in}=0\,.
\]
Obviously, from $J_{\pm1}^{\pm}(x)$ we obtain the energy momentum tensor.

\textbf{Examples:} Up to normalization constants we calculate for the order
parameters $\sigma_{1}(x)$ and $\sigma_{2}(x)$
\begin{align*}
\langle\,0\,|\,\sigma_{1}(0)\,|\,\theta\rangle_{1}  &  =1\\
\langle\,0\,|\,\sigma_{2}(0)\,|\,\theta_{1},\theta_{2}\rangle_{11}^{in}  &
=\frac{F(\theta_{12})}{\sinh\frac{1}{2}(\theta_{12}-2\pi i/N)\sinh\frac{1}%
{2}(\theta_{12}+2\pi i/N)}%
\end{align*}
and for the para-fermi fields $\,\psi_{Q}(x)$ and $\psi_{2}(x)$%
\begin{align}
\langle\,0\,|\,\psi_{Q}(0)\,|\,\theta\rangle_{Q}  &  =e^{\frac{Q(N-Q)}%
{N}\theta}\label{4.4}\\
\langle\,0\,|\,\psi_{2}(0)\,|\,\theta_{1},\theta_{2}\rangle_{11}^{in}  &
=\frac{e^{\left(  1-\frac{2}{N}\right)  \left(  \theta_{1}+\theta_{2}\right)
}F(\theta_{12})}{\sinh\frac{1}{2}(\theta_{12}-2\pi i/N)\sinh\frac{1}{2}%
(\theta_{12}+2\pi i/N)}\nonumber
\end{align}
where $|\,\theta\rangle_{Q}$ denotes a one-particle state of charge $Q$ and
$|\,\theta_{1},\theta_{2}\rangle_{11}^{in}$ a state of two particles of charge
1.

\section{Commutation rules\label{s5}}

\subsection{The general formula}

Techniques similar to the ones of this section have been applied for the
simpler case of no bound states and bosonic statistics in \cite{Sm,Las}. A
generalization for the case of bound states has been discussed in \cite{Q}.
Here we generalize these techniques for the case of more general statistics
and also discuss the contribution of poles related to the double poles of
bound state S-matrices\footnote{For bound state form factors there are also
higher order 'physical poles' (see e.g. \cite{CT,BCDS,DM,AMV}).}. In order to
discuss commutation rules of two fields $\phi(x)$ and $\psi(y)$ we have to use
a general crossing formula for form factors which was derived in \cite{BK}
(see also \cite{Sm}). For quantum field theories with general statistics we
introduce assumptions on the statistics factor of a field $\psi(x)$ and a
particle $\alpha$. It is easy to see that for consistency of (ii) and (iii)
the condition $\sigma^{\psi}(\alpha)\sigma^{\psi}(\bar{\alpha})=1$ has to hold
if $\bar{\alpha}$ is the anti-particle of $\alpha$. We assume that
\begin{equation}
\sigma^{\psi}(\alpha)=\sigma^{\psi}(Q_{\alpha}) \label{5.1}%
\end{equation}
depends on the charge of the particle such that $\sigma^{\psi}(Q+Q^{\prime
})=\sigma^{\psi}(Q)\sigma^{\psi}(Q^{\prime})$. A stronger assumption (which
holds for the $Z(N)$-model) is the existence of a `dual charge' $\tilde
{Q}_{\psi}$ of the fields such that%
\begin{equation}
\sigma^{\psi}(\alpha)=\omega^{\tilde{Q}_{\psi}Q_{\alpha}} \label{5.3}%
\end{equation}
where $\left\vert \omega\right\vert =1$.

In order to write the following long formulae we introduce short notation: For
a field $\mathcal{O}(x)$ and for ordered sets of rapidities $\theta_{1}%
>\dots>\theta_{n}$ and $\theta_{1}^{\prime}<\dots<\theta_{m}^{\prime}$ we
write the general matrix element of $\mathcal{O}(0)$ as%
\begin{equation}
\mathcal{O}_{\underline{\alpha}}^{\underline{\beta}}(\underline{\theta}%
_{\beta}^{\prime},\underline{\theta}_{\alpha}):=\,^{~out}\langle\beta
_{m}(\theta_{m}^{\prime}),\dots,\beta_{1}(\theta p_{1}^{\prime}%
)\,|\,\mathcal{O}\,|\,\alpha_{1}(\theta_{1}),\dots,\alpha_{n}(\theta
_{n})\,\rangle^{in} \label{5.7}%
\end{equation}
where $\underline{\theta}_{\alpha}=(\theta_{1},\dots,\theta_{n})$ and
$\underline{\theta}_{\beta}^{\prime}=(\theta_{1}^{\prime},\dots,\theta
_{m}^{\prime})$. The array of indices $\underline{\alpha}=(\alpha_{1}%
,\dots,\alpha_{n})$ denote a set of particles ($\alpha_{i}\in\left\{
\text{types of particles}\right\}  $) and correspondingly for $\underline
{\beta}$ (we also write $\left\vert \alpha\right\vert =n$ etc.). Similar as
for form factors this matrix element is given for general order of the
rapidities by the symmetry property (i) for both the $in$- and $out$-states
which takes the general form:
\[
\mathcal{O}_{\underline{\alpha}}^{\underline{\beta}}(\underline{\theta}%
_{\beta}^{\prime},\underline{\theta}_{\alpha})=S_{\underline{\delta}%
}^{\underline{\beta}}(\underline{\theta}_{\delta}^{\prime})\mathcal{O}%
_{\underline{\gamma}}^{\underline{\delta}}(\underline{\theta}_{\delta}%
^{\prime},\underline{\theta}_{\gamma})S_{\underline{\alpha}}^{\underline
{\gamma}}(\underline{\theta}_{\alpha})
\]
if $\underline{\theta}_{\gamma}$ is a permutation of $\underline{\theta
}_{\alpha}$ and $\underline{\theta}_{\delta}^{\prime}$ a permutation of
$\underline{\theta}_{\beta}^{\prime}$. The matrix $S_{\underline{\alpha}%
}^{\underline{\gamma}}(\underline{\theta}_{\alpha})$ is defined as the
representation of the permutation $\pi(\underline{\theta}_{\alpha}%
)=\underline{\theta}_{\gamma}$ generated by the two-particle S-matrices
$S_{\alpha_{1}\alpha_{2}}^{\gamma_{2}\gamma_{1}}(\theta_{12})$, for example
$S_{\alpha_{1}\alpha_{2}\alpha_{3}}^{\gamma_{3}\gamma_{2}\gamma_{1}}%
(\theta_{1},\theta_{2},\theta_{3})=S_{\alpha_{1}\lambda_{3}}^{\gamma_{3}%
\gamma_{1}}(\theta_{13})S_{\alpha_{2}\alpha_{3}}^{\lambda_{3}\gamma_{2}%
}(\theta_{23})$ (c.f. \cite{BK}).

We consider an arbitrary matrix element of products of fields
$\mathcal{O}=\phi(x)\psi(y)$ and $\mathcal{O}=\psi(y)\phi(x)$. Inserting a
complete set of intermediate states $|\,\underline{\tilde{\theta}}_{\gamma
}\,\rangle_{\underline{\gamma}}^{in}$ we obtain%
\begin{equation}
\left(  \phi(x)\psi(y)\right)  _{\underline{\alpha}}^{\underline{\beta}%
}(\underline{\theta}_{\beta}^{\prime},\underline{\theta}_{\alpha
})=e^{iP_{\beta}^{\prime}x-iP_{\alpha}y}\frac{1}{\gamma!}\int_{\underline
{\tilde{\theta}}_{\gamma}}\phi_{\underline{\gamma}}^{\underline{\beta}%
}(\underline{\theta}_{\beta}^{\prime},\underline{\tilde{\theta}}_{\gamma}%
)\psi_{\underline{\alpha}}^{\underline{\gamma}}(\underline{\tilde{\theta}%
}_{\gamma},\underline{\theta}_{\alpha})e^{-i\tilde{P}_{\gamma}(x-y)}
\label{5.9}%
\end{equation}
where $\phi=\phi(0),\psi=\psi(0),$ $P_{\alpha}=$ the total momentum of the
state $|\,\underline{\theta}_{\alpha}\,\rangle_{\underline{\alpha}}^{in}$ etc.
and $\int_{\underline{\tilde{\theta}}_{\gamma}}=\prod_{k=1}^{|\gamma|}%
\int\frac{d\tilde{\theta}_{k}}{4\pi}$. Einstein summation convention over all
sets $\underline{\gamma}$ is assumed. We also define $\gamma!=\prod_{\alpha
}n_{\alpha}!$ where $n_{\alpha}$ is the number of particles of type $\alpha$
in $\underline{\gamma}$. We apply the general crossing formula (31) of
\cite{BK} which is obtained by taking into account the disconnected terms in
(ii) and iterating that formula. Strictly speaking, we apply the second
version of the crossing formula to the matrix element of $\phi$
\begin{equation}
\phi_{\underline{\gamma}}^{\underline{\beta}}(\underline{\theta}_{\beta
}^{\prime},\underline{\tilde{\theta}}_{\gamma})=\sum_{\substack{\underline
{\theta}_{\rho}^{\prime}\cup\underline{\theta}_{\tau}^{\prime}=\underline
{\theta}_{\beta}^{\prime} \\\underline{\tilde{\theta}}_{\varsigma}%
\cup\underline{\tilde{\theta}}_{\varkappa}=\underline{\tilde{\theta}}_{\gamma
}}}S_{\underline{\rho}\underline{\tau}}^{\underline{\beta}}(\underline{\theta
}_{\rho}^{\prime},\underline{\theta}_{\tau}^{\prime})\,\phi_{\underline
{\varsigma}\underline{\bar{\rho}}}(\underline{\tilde{\theta}}_{\varsigma
},\underline{\theta^{\prime}}_{\bar{\rho}}-i\pi_{-})\mathbf{C}^{\underline
{\rho}\underline{\bar{\rho}}}\,\mathbf{1}_{\underline{\sigma}}^{\underline
{\tau}}(\underline{\theta^{\prime}}_{\tau},\underline{\tilde{\theta}%
}_{\varkappa})\,S_{\underline{\gamma}}^{\underline{\varsigma}\underline
{\sigma}}(\underline{\tilde{\theta}}_{\gamma}) \label{5.93}%
\end{equation}
where $\underline{\bar{\rho}}=\left(  \bar{\rho}_{|\rho|},\dots,\bar{\rho}%
_{1}\right)  $ with $\bar{\rho}=$ antiparticle of $\rho$ and $\underline
{\theta^{\prime}}_{\bar{\rho}}-i\pi_{-}$ means that all rapidities are taken
as $\theta^{\prime}-i\left(  \pi-\epsilon\right)  $. The matrix $\mathbf{1}%
_{\underline{\sigma}}^{\underline{\tau}}(\underline{\theta^{\prime}}_{\tau
},\underline{\tilde{\theta}}_{\varkappa})$ is defined by (\ref{5.7}) with
$\mathcal{O}=\mathbf{1}$ the unit operator. The summation is over all
decompositions of the sets of rapidities $\underline{\theta}_{\beta}^{\prime}
$ and $\underline{\tilde{\theta}}_{\gamma}$. To the matrix element of $\psi$
we apply the first version of the crossing formula
\begin{equation}
\psi_{\underline{\alpha}}^{\underline{\gamma}}(\underline{\tilde{\theta}%
}_{\gamma},\underline{\theta}_{\alpha})=\sigma_{(\underline{\bar{\gamma}}%
)}^{\psi}\sum_{\substack{\underline{\tilde{\theta}}_{\nu}\cup\underline
{\tilde{\theta}}_{\pi}=\underline{\tilde{\theta}}_{\gamma} \\\underline
{\theta}_{\mu}\cup\underline{\theta}_{\lambda}=\underline{\theta}_{\alpha}%
}}S_{\underline{\nu}\underline{\pi}}^{\underline{\gamma}}(\underline
{\tilde{\theta}}_{\nu},\underline{\tilde{\theta}}_{\pi})\,\mathbf{1}%
_{\underline{\mu}}^{\underline{\nu}}(\underline{\tilde{\theta}}_{\nu
},\underline{\theta}_{\mu})\,\mathbf{C}^{\underline{\pi}\underline{\bar{\pi}}%
}\psi_{\underline{\bar{\pi}}\underline{\lambda}}(\underline{\tilde{\theta}%
}_{\bar{\pi}}+i\pi_{-},\underline{\theta}_{\lambda})\,S_{\underline{\gamma}%
}^{\underline{\mu}\underline{\lambda}}(\underline{\theta}_{\gamma})
\label{5.95}%
\end{equation}
where we assume that the statistics factor $\sigma_{(\underline{\bar{\gamma}%
})}^{\psi}$ of the field $\psi$ with respect to all particles in
$\underline{\gamma}$ is the same for all $\underline{\gamma}$ which contribute
to (\ref{5.9}) (see below). Inserting (\ref{5.93}) and (\ref{5.95}) in
(\ref{5.9}) we use the product formula $S_{\underline{\gamma}}^{\underline
{\varsigma}\underline{\sigma}}(\underline{\tilde{\theta}}_{\gamma
})S_{\underline{\nu}\underline{\pi}}^{\underline{\gamma}}(\underline
{\tilde{\theta}}_{\nu},\underline{\tilde{\theta}}_{\pi})=S_{\underline{\nu
}\underline{\pi}}^{\underline{\varsigma}\underline{\sigma}}(\underline
{\tilde{\theta}}_{\nu},\underline{\tilde{\theta}}_{\pi})$. Let us first assume
that the sets rapidities in the initial state $\underline{\theta}_{\alpha}$
and the ones of the final state $\underline{\theta}_{\beta}^{\prime}$ have no
common elements which implies that also $\underline{\tilde{\theta}}_{\nu}%
\cap\underline{\tilde{\theta}}_{\sigma}=\emptyset$. Then we may use (ii) to
get $S_{\underline{\nu}\underline{\pi}}^{\underline{\varsigma}\underline
{\sigma}}(\underline{\tilde{\theta}}_{\nu},\underline{\tilde{\theta}}_{\pi
})=1$ and we can perform the $\underline{\tilde{\theta}}_{\nu}$- and
$\underline{\tilde{\theta}}_{\varkappa}$-integrations. The remaining
$\tilde{\theta}$-integration variables are $\underline{\tilde{\theta}}%
_{\omega}=\underline{\tilde{\theta}}_{\varsigma}\cap\underline{\tilde{\theta}%
}_{\pi}$, then we may write for the sets of particles $\underline{\varsigma
}=\underline{\mu}\underline{\omega},\,\underline{\pi}=\underline{\omega
}\underline{\tau}$ and $\underline{\gamma}=\underline{\mu}\underline{\omega
}\underline{\tau}$ and similar for rapidities and momenta. Equation
(\ref{5.9}) simplifies as%
\begin{multline}
\left(  \phi(x)\psi(y)\right)  _{\underline{\alpha}}^{\underline{\beta}%
}(\underline{\theta}_{\beta}^{\prime},\underline{\theta}_{\alpha}%
)=\sum_{\substack{\underline{\theta}_{\rho}^{\prime}\cup\underline{\theta
}_{\tau}^{\prime}=\underline{\theta}_{\beta}^{\prime} \\\underline{\theta
}_{\mu}\cup\underline{\theta}_{\lambda}=\underline{\theta}_{\alpha}}}\frac
{\mu!\tau!}{\mu\omega\tau!}S_{\underline{\rho}\underline{\tau}}^{\underline
{\beta}}(\underline{\theta}_{\rho}^{\prime},\underline{\theta}_{\tau}^{\prime
})\int_{\underline{\tilde{\theta}}_{\omega}}X_{\underline{\mu}\underline
{\lambda}}^{\underline{\rho}\underline{\tau}}\label{phipsi}\\
\times S_{\underline{\alpha}}^{\underline{\mu}\underline{\lambda}}%
(\underline{\theta}_{\alpha})e^{i\left(  P_{\rho}^{\prime}-P_{\mu}\right)
x-i\left(  P_{\lambda}-P_{\tau}^{\prime}\right)  y}%
\end{multline}
where%
\begin{multline}
X_{\underline{\mu}\underline{\lambda}}^{\underline{\rho}\underline{\tau}%
}=\sigma_{(\underline{\bar{\gamma}})}^{\psi}\phi_{\underline{\mu}%
\underline{\omega}\underline{\bar{\rho}}}(\underline{\theta}_{\mu}%
,\underline{\tilde{\theta}}_{\omega},\underline{\theta^{\prime}}_{\bar{\rho}%
}-i\pi_{-})\mathbf{C}^{\underline{\bar{\rho}}\underline{\rho}}\mathbf{C}%
^{\underline{\tau}\underline{\bar{\tau}}}\mathbf{C}^{\underline{\omega
}\underline{\bar{\omega}}}\label{X}\\
\times\psi_{\underline{\bar{\tau}}\underline{\bar{\omega}}\underline{\lambda}%
}(\underline{\theta}_{\bar{\tau}}^{\prime}+i\pi_{-},\underline{\tilde{\theta}%
}_{\bar{\omega}}+i\pi_{-},\underline{\theta}_{\lambda})e^{-i\tilde{P}_{\omega
}(x-y)}\,.
\end{multline}
The integrand $X_{\underline{\mu}\underline{\lambda}}^{\underline{\rho
}\underline{\tau}}$ may be depicted as%
\[
X_{\underline{\mu}\underline{\lambda}}^{\underline{\rho}\underline{\tau}%
}~=~\sigma_{(\underline{\bar{\gamma}})}^{\psi}~~%
\begin{array}
[c]{c}%
\unitlength4mm\begin{picture}(11,6) \put(2,4){\oval(4,2)} \put(1.6,3.7){$\phi$}
\put(1,0){\line(0,1){3}} \put(0,.4){$\underline{\mu}$}
\put(4,3){\oval(2,2)[b]} \put(5,3){\line(0,1){3}}
\put(4.1,5.3){$\underline{\rho}$} \put(5.5,3){\oval(7,4)[b]}
\put(1.4,1.1){$\underline{\omega}$} \put(9,1.1){$\underline{\bar\omega}$}
\put(9,4){\oval(4,2)} \put(8.6,3.7){$\psi$} \put(7,3){\oval(2,2)[b]}
\put(6,3){\line(0,1){3}} \put(6.3,5.3){$\underline{\tau}$}
\put(10,0){\line(0,1){3}} \put(10.3,.4){$\underline{\lambda}$} \end{picture}
\end{array}
\]
Similarly, if we apply for the operator product $\psi(y)\phi(x)$ again the
second crossing formula to the matrix element of $\phi$ and the first one the
matrix element of $\psi$ we obtain equation (\ref{phipsi}) where
$X_{\underline{\mu}\underline{\lambda}}^{\underline{\rho}\underline{\tau}}$ is
replaced by%
\begin{multline}
Y_{\underline{\mu}\underline{\lambda}}^{\underline{\rho}\underline{\tau}%
}=\sigma_{(\underline{\bar{\beta}})}^{\psi}\phi_{\underline{\mu}%
\underline{\omega}\underline{\bar{\rho}}}(\underline{\theta}_{\mu}%
,\underline{\tilde{\theta}}_{\omega}-i\pi_{-},\underline{\theta^{\prime}%
}_{\bar{\rho}}-i\pi_{-})\mathbf{C}^{\underline{\bar{\rho}}\underline{\rho}%
}\mathbf{C}^{\underline{\tau}\underline{\bar{\tau}}}\mathbf{C}^{\underline
{\omega}\underline{\bar{\omega}}}\label{Y}\\
\times\psi_{\underline{\bar{\tau}}\underline{\bar{\omega}}\underline{\lambda}%
}(\underline{\theta}_{\bar{\tau}}^{\prime}+i\pi_{-},\underline{\tilde{\theta}%
}_{\bar{\omega}},\underline{\theta}_{\lambda})e^{i\tilde{P}_{\omega}(x-y)}%
\end{multline}
which means that only $\sigma_{(\underline{\bar{\gamma}})}^{\psi}$ is
replaced by
$\sigma_{(\underline{\bar{\beta}})}^{\psi},\,\tilde{P}_{\omega}$ by
$-\tilde{P}_{\omega}$ and the integration variables $\underline{\tilde{\theta
}}_{\omega}$ by $\underline{\tilde{\theta}}_{\bar{\omega}}-i\pi_{-}$.

\textbf{No bound states:} In this case there are no singularities in the
physical strip and we may shift in the matrix element of $\psi(y)\phi(x)$
(\ref{phipsi}) with (\ref{Y}) for equal times and $x^{1}<y^{1}$ the
integration variables by $\tilde{\theta}_{i}\rightarrow\tilde{\theta}_{i}%
+i\pi_{-}$. Note that the factor $e^{i\tilde{P}_{\omega}(x-y)}$ decreases for
$0<\operatorname{Re}\tilde{\theta}_{i}<\pi$ if $x^{1}<y^{1}$. Because
$\tilde{P}_{\omega}\rightarrow-\tilde{P}_{\omega}$ we get the matrix element
of $\phi(x)\psi(y)$ (\ref{phipsi}) with (\ref{X}) up to the statistics
factors. Therefore we conclude
\begin{equation}
\phi(x)\psi(y)=\psi(y)\phi(x)\sigma^{\psi\phi}\quad\text{for }x^{1}<y^{1}
\label{5.11}%
\end{equation}
where $\sigma^{\psi\phi}=\sigma_{(\underline{\bar{\gamma}})}^{\psi}%
/\sigma_{(\underline{\bar{\beta}})}^{\psi}$. Using the assumption (\ref{5.1})
we have with $Q_{\underline{\bar{\gamma}}}=\sum_{\gamma\in\underline{\gamma}%
}Q_{\bar{\gamma}}$%
\[
\sigma_{(\underline{\bar{\gamma}})}^{\psi}=\prod_{\gamma\in\underline{\gamma}%
}\sigma^{\psi}(\bar{\gamma})=\sigma^{\psi}(Q_{\underline{\bar{\gamma}}%
})=\sigma^{\psi}(Q_{\underline{\bar{\beta}}}-Q_{\phi})
\]
which is the same for all $\underline{\gamma}$, as assumed above. The last
equation follows from $Q_{\bar{\gamma}}=-Q_{\gamma}$ and charge conservation
which means that the the matrix elements $\phi_{\underline{\gamma}%
}^{\underline{\beta}}$ in (\ref{5.9}) are non-vanishing if $Q_{\underline
{\beta}}+Q_{\phi}=Q_{\underline{\gamma}}$. Therefore the statistics factor of
the fields $\psi$ with respect to $\phi$ is
\begin{equation}
\sigma^{\psi\phi}=\frac{\sigma^{\psi}(Q_{\underline{\bar{\gamma}}})}%
{\sigma^{\psi}(Q_{\underline{\bar{\beta}}})}=\sigma^{\psi}(-Q_{\phi}%
)=1/\sigma^{\psi}(Q_{\phi}) \label{10}%
\end{equation}
which is in general not symmetric under the exchange of $\psi$ and $\phi$.
Finally, we obtain the space like commutation rules%
\begin{equation}
\phi(x)\psi(y)=\psi(y)\phi(x)\left\{
\begin{array}
[c]{lll}%
1/\sigma^{\psi}(Q_{\phi}) & \text{for} & x^{1}<y^{1}\\
\sigma^{\phi}(Q_{\psi}) & \text{for} & x^{1}>y^{1}%
\end{array}
\right.  \label{12}%
\end{equation}
where the second relation is obtained from (\ref{5.11}) by exchanging
$\phi\leftrightarrow\psi$ and $x\leftrightarrow y$ . The same result appears
when there are bound states. This is proved in appendix \ref{a3} where also
the existence of double poles in bound state S-matrices is taken into account.

\subsection{Application to the $Z(N)$-model}

The statistics factors in this model are of the form (\ref{5.3}) $\sigma
^{\psi}(\alpha)=\omega^{\tilde{Q}_{\psi}Q_{\alpha}}$ where $\tilde{Q}_{\psi}$
is the dual charge of the field $\psi$ and $Q_{\alpha}$ is the charge of the
particle $\alpha$, therefore $\sigma^{\psi\phi}=\omega^{-\tilde{Q}_{\psi
}Q_{\phi}}$. The general equal time commutation rule (\ref{12}) for fields
$\psi_{Q\tilde{Q}}(x)$ defined by (\ref{4.2}) in section \ref{s4} reads as%
\begin{equation}
\psi_{Q\tilde{Q}}(x)\psi_{R\tilde{R}}(y)=\psi_{R\tilde{R}}(y)\psi_{Q\tilde{Q}%
}(x)\left\{
\begin{array}
[c]{lll}%
\omega^{-\tilde{R}Q}=e^{-2\pi i\tilde{R}Q/N} & \text{for} & x^{1}<y^{1}\\
\omega^{\tilde{Q}R}=e^{2\pi i\tilde{Q}R/N} & \text{for} & x^{1}>y^{1}\,.
\end{array}
\right.  \label{13}%
\end{equation}
Notice that in this model we have a more general anyonic statistics.

\noindent\textbf{Examples:}

\begin{enumerate}
\item The order parameters have bosonic commutation rules with respect to each
other%
\[
\sigma_{Q}(x)\sigma_{Q^{\prime}}(y)=\sigma_{Q^{\prime}}(y)\sigma_{Q}(x)\,.
\]

\item The disorder parameters have again bosonic commutation rules with
respect to each other.

\item For the order-disorder parameters we obtain the typical commutation
rule
\[
\mu_{\tilde{Q}}(x)\sigma_{Q}(y)=\sigma_{Q}(y)\mu_{\tilde{Q}}(x)\left\{
\begin{array}
[c]{lll}%
1 & \text{for} & x^{1}<y^{1}\\
\omega^{\tilde{Q}Q}=e^{2\pi i\tilde{Q}Q/N} & \text{for} & x^{1}>y^{1}\,.
\end{array}
\right.
\]

\item The para-fermi fields have anyonic commutation rules%
\begin{equation}
\psi_{Q}(x)\psi_{R}(y)=\psi_{R}(y)\psi_{Q}(x)e^{\epsilon(x^{1}-y^{1})2\pi
iQR/N}\,. \label{pfer}%
\end{equation}

\end{enumerate}

\noindent These results prove the commutation rules (\ref{cr}) in the
Introduction.

\noindent\textbf{The 2-point Wightman function:} In order to compare these
commutation rules with the explicit results of the previous section we
calculate the 2-point Wightman function for the para-fermi fields $\psi_{Q}$
and $\psi_{N-Q}$ (with spin $s=Q(N-Q)/N$) in 1-particle (charge $Q$)
intermediate state approximation. Using the result (\ref{4.4}) we obtain
\begin{multline*}
\langle\,0\,|\,\psi_{Q}(x)\,\psi_{N-Q}(0)|\,0\rangle=\int\frac{d\theta}{4\pi
}\langle\,0\,|\,\psi_{Q}(x)|\,\theta\rangle_{Q}\,_{Q}\langle\,\theta
|\psi_{N-Q}(0)|\,0\rangle+\dots\\
=\frac{1}{2\pi}\left(  \frac{x^{-}-i\epsilon}{x^{+}-i\epsilon}\right)
^{\nu/2}\operatorname{K}_{\nu}\left(  M\sqrt{i\left(  x^{+}-i\epsilon\right)
}\sqrt{i\left(  x^{-}-i\epsilon\right)  }\right)  +\dots
\end{multline*}
where $\nu=2Q(N-Q)/N$ and $x^{\pm}=t\mp x$. This agrees with the commutation
rule (\ref{pfer}), because for $t=0$ and $x>0$ using the symmetry
$Q\leftrightarrow N-Q,x\rightarrow-x$ and translation invariance we obtain
\begin{align*}
\langle\,0\,|\,\psi_{Q}(x)\,\psi_{N-Q}(0)|\,0\rangle &  =\langle
\,0\,|\,\psi_{N-Q}(x)\,\psi_{Q}(0)|\,0\rangle\\
&  =e^{i\pi\nu}\langle\,0\,|\,\psi_{N-Q}(-x)\,\psi_{Q}(0)|\,0\rangle\\
&  =e^{i\pi\nu}\langle\,0\,|\,\psi_{N-Q}(0)\,\psi_{Q}(x)|\,0\rangle
\end{align*}
where $\left(  \left(  x-i\epsilon\right)  /\left(  -x-i\epsilon\right)
\right)  ^{\nu/2}=e^{i\pi\nu\epsilon(x)/2}$ has bee used. The asymptotic
behavior is obtained from
\[
2K_{\nu}\left(  z\right)  \rightarrow\left\{
\begin{array}
[c]{lll}%
\Gamma(\nu)\left(  \tfrac{z}{2}\right)  ^{-\nu}+\Gamma(-\nu)\left(  \tfrac
{z}{2}\right)  ^{\nu} & \text{for} & z\rightarrow0\\[2mm]%
\sqrt{\tfrac{2\pi}{z}}e^{-z} & \text{for} & z\rightarrow\infty
\end{array}
\right.
\]
for $\nu\neq0$. Therefore the leading short distance behavior is up to
constants%
\begin{align*}
\langle\,0\,|\,\psi_{Q}(x)\,\psi_{N-Q}(0)|\,0\rangle &  \thicksim\left(
x^{+}-i\epsilon\right)  ^{-\nu}\\
\langle\,0\,|\,\tilde{\psi}_{Q}(x)\,\tilde{\psi}_{N-Q}(0)|\,0\rangle &
\thicksim\left(  x^{-}-i\epsilon\right)  ^{-\nu}%
\end{align*}
where the fields $\tilde{\psi}_{Q}(x)$ are obtained by changing the sign in
the exponent of (\ref{4.2}).

\section*{Acknowledgments}

We thank V.A. Fateev, R. Flume, A. Fring, A. Nersesyan, R. Schrader, B.
Schroer, J. Teschner, A. Tsvelik, Al.B. Zamolodchikov and A.B. Zamolodchikov
for discussions. In particular we thank V.A. Fateev for bringing the preprint
\cite{KiSm} to our attention and F.A. Smirnov for sending a copy. H.B. was
supported by DFG, Sonderforschungsbereich 288 `Differentialgeometrie und
Quantenphysik', partially by the grants INTAS 99-01459 and INTAS 00-561 and in
part by Volkswagenstiftung within in the project "Nonperturbative aspects of
quantum field theory in various space-time dimensions". A. F. acknowledges
support from PRONEX under contract CNPq 66.2002/1998-99 and CNPq (Conselho
Nacional de Desenvolvimento Cient\'{\i}fico e Tecnol\'{o}gico). This work is
also supported by the EU network EUCLID, 'Integrable models and applications:
from strings to condensed matter', HPRN-CT-2002-00325.

\appendix

\section*{Appendix}

\section{Some useful formulae \label{a1}}

In this appendix we provide some explicit formulae for the scattering matrices
and the minimal form factors which we frequently employ in the explicit
computations. The S-matrix of two 'fundamental' particles (i.e. of type 1) is
\cite{KS}
\[
S(\theta)=\frac{\sinh\frac{1}{2}(\theta+\frac{2\pi i}{N})}{\sinh\frac{1}%
{2}(\theta-\frac{2\pi i}{N})}=-\exp\int_{0}^{\infty}\frac{dt}{t}\,2\frac{\sinh
t(1-2/N)}{\sinh t}\sinh tx\,
\]
where $\theta$ is the rapidity difference defined by
\[
p_{1}p_{2}=m^{2}\cosh\theta\,.
\]
A particle of type $\alpha\;(0<\alpha<N)$ is a bound state $\alpha=(\alpha
_{1}\cdots\alpha_{l})$ of particles of type $\alpha_{i}$ where $\alpha
=\alpha_{1}+\cdots+\alpha_{l}$, in particular $\alpha=(\underbrace{1\cdots
1}_{\alpha})$ for all $\alpha_{i}=1$. For the scattering of the bound state
$\alpha$ and $\beta$ we have \cite{K3}
\[
S_{\alpha\beta}(\theta)=\exp2\int_{0}^{\infty}\frac{dx}{x}\frac{\cosh
x(1-\frac{|\beta-\alpha|}{N})-\cosh x(1-\frac{\beta+\alpha}{N})}{\sinh
x\tanh(x/N)}\sinh x\frac{\theta}{i\pi}\,.
\]
The minimal form factor functions, which satisfies Watson's equations, are
obtained from the S-matrix formulae \cite{KW} and are given as ($\beta>\alpha
$)
\[
F_{\alpha\beta}^{\min}(\theta)=\exp\int_{0}^{\infty}\frac{dt}{t}\,2\frac{\sinh
t\left(  1-\frac{\beta}{N}\right)  \sinh t\left(  \frac{\alpha}{N}\right)
}{\sinh^{2}t\tanh t/N}\left(  1-\cosh t\left(  1-\frac{\theta}{i\pi}\right)
\right)
\]
in particular\cite{K3}
\begin{multline*}
F_{11}^{\min}(\theta)=\exp\int_{0}^{\infty}\frac{dt}{t}\,2\frac{\sinh t\left(
1-\frac{1}{N}\right)  \cosh t\frac{1}{N}}{\sinh^{2}t}\left(  1-\cosh t\left(
1-\frac{\theta}{i\pi}\right)  \right) \\
=-i\sinh\tfrac{1}{2}\theta\exp\int_{0}^{\infty}\frac{dt}{t}\,\frac{\sinh
t\left(  1-\frac{2}{N}\right)  }{\sinh^{2}t}\left(  1-\cosh t\left(
1-\frac{\theta}{i\pi}\right)  \right) \\
=-i\sinh\tfrac{1}{2}\theta\prod_{k=0}^{\infty}\frac{\Gamma(k+1-\frac{1}%
{N}+\frac{x}{2})}{\Gamma(k+\frac{1}{N}+\frac{x}{2})}\,\frac{\Gamma
(k+2-\frac{1}{N}-\frac{x}{2})}{\Gamma(k+1+\frac{1}{N}-\frac{x}{2})}\left(
\frac{\Gamma(k+\frac{1}{2}+\frac{1}{N})}{\Gamma(k+\frac{3}{2}-\frac{1}{N}%
)}\right)  ^{2}%
\end{multline*}
and with $\bar{1}=(N-1)$%
\begin{align*}
F_{1\bar{1}}^{\min}(\theta)  &  =\exp\int_{0}^{\infty}\frac{dt}{t}%
\,\frac{\sinh t\frac{2}{N}}{\sinh^{2}t}\left(  1-\cosh t\left(  1-\frac
{\theta}{i\pi}\right)  \right) \\
&  =\prod_{k=0}^{\infty}\frac{\Gamma(k+\frac{1}{2}+\frac{1}{N}+\frac{\theta
}{2\pi i})}{\Gamma(k+\frac{1}{2}-\frac{1}{N}+\frac{\theta}{2\pi i})}%
\,\frac{\Gamma(k+\frac{3}{2}+\frac{1}{N}-\frac{\theta}{2\pi i})}%
{\Gamma(k+\frac{3}{2}-\frac{1}{N}-\frac{\theta}{2\pi i})}\left(  \frac
{\Gamma(k+1-\frac{1}{N})}{\Gamma(k+1+\frac{1}{N})}\right)  ^{2}.
\end{align*}
There are simple relations between the minimal form factors which we
essentially use in our construction which are up to constants%
\begin{gather*}
F_{11}^{\min}\left(  \theta+\frac{i\pi}{N}\right)  F_{11}^{\min}\left(
\theta-\frac{i\pi}{N}\right)  \propto\sinh\tfrac{1}{2}\left(  \theta
+\frac{i\pi}{N}\right)  \sinh\tfrac{1}{2}\left(  \theta-\frac{i\pi}{N}\right)
F_{12}^{\min}(\theta)\\
\prod_{k=0}^{N-1}F_{11}^{\min}(\theta+\frac{k}{N}2\pi i)\propto\prod
_{k=0}^{N-2}\sinh\tfrac{1}{2}\left(  \theta+\frac{k}{N}2\pi i\right)
\sinh\tfrac{1}{2}\left(  \theta+\frac{k+1}{N}2\pi i\right) \\
F_{11}^{\min}(\theta)F_{1\bar{1}}^{\min}(\theta+i\pi)\propto\sinh\tfrac{1}%
{2}\theta\sinh\tfrac{1}{2}\left(  \theta+2i\pi/N\right)  \,.
\end{gather*}
In equations (\ref{F3}) and (\ref{FN}) we used the function $F(\theta
)=c_{N}F_{11}^{\min}(\theta)$ with
\begin{equation}
c_{N}=e^{i\pi\frac{N-1}{N}}\exp\left(  \int_{0}^{\infty}\frac{dt}{t\sinh
t}\,\left(  \left(  1-\frac{2}{N}\right)  -\frac{\sinh t\left(  1-\frac{2}%
{N}\right)  }{\sinh t}\right)  \right)  \label{c}%
\end{equation}
such that the normalizations in (\ref{phiF}) and (\ref{phi}) hold.

\section{Integrals for the $Z(2)$-model\label{ab}}

The claim (\ref{2.4}) follows from the following lemma

\begin{lemma}
For $n=2m+1$ odd and $x_{i}=e^{\theta_{i}}$%
\[
f_{n}(\underline{x}):=I_{nm}(\underline{\theta},1)-\left(  2i\right)
^{(n-1)/2}\prod_{1\leq i<j\leq n}\frac{\tanh\tfrac{1}{2}\theta_{ij}}%
{F(\theta_{ij})}=0\,.
\]

\end{lemma}

\begin{proof}
Again as in the proof of lemma 2 in \cite{BK1} we apply induction and
Liouville's theorem. One easily verifies $f_{1}(\underline{x})=f_{3}%
(\underline{x})=0$. As induction assumptions we take $f_{n-2}=0$. The
functions $f_{n}(\underline{x})$ are a meromorphic functions in terms of the
$x_{i}$ with at most simple poles at $x_{i}=-x_{j}$ since pinchings appear for
$z_{k}=\theta_{i}=\theta_{j}\pm i\pi$. The residues of the poles are
proportional to $f_{n-2}$ as follows from the recursion relations (iii) for
both terms. Furthermore $f_{n}(\underline{x})\rightarrow0$ for $x_{i}%
\rightarrow\infty$. Therefore $f_{n}(\underline{x})$ vanishes identically by
Liouville's theorem.
\end{proof}

Note that the integrations in the definition (\ref{Inm2}) of $I_{nm}$ can
easily be performed and with $\mathcal{N}=\{1,\dots,n\}$ and $|\mathcal{K}|=m$%
\[
I_{nm}(\underline{\theta},1)=\sum_{\mathcal{K}\subset\mathcal{N}}\prod
_{k\in\mathcal{K}}\prod_{i\in\mathcal{N}\setminus\mathcal{K}}\frac{2i}%
{\sinh\theta_{ki}}\,.
\]

\section{ Proof of the main lemma \label{a2}}

In this appendix we prove the main lemma \ref{l4} which provides the general
$Z(N)$-form factor formula.

\begin{proof}
Similar as in the proof of lemma \ref{l3} we calculate%
\begin{multline*}
\operatorname*{Res}_{\theta_{N-1N}=i\eta}\cdots\operatorname*{Res}%
_{\theta_{12}=i\eta}I_{n\underline{m}}(\underline{\theta},p_{n\underline{m}%
}^{{\mathcal{O}}})=\frac{m_{1}\dots m_{N-1}}{m_{1}!\dots m_{N-1}!}\left(
\prod_{k=1}^{N-1}\prod_{j=2}^{m_{k}}\int_{\mathcal{C}_{\underline{\theta
}^{\prime}}}\frac{dz_{kj}}{R}\right) \\
\times\prod_{k=1}^{N-1}\left(  \prod_{i=N+1}^{n}\prod_{j=2}^{m_{k}}\phi
(z_{kj}-\theta_{i})\prod_{2\leq i<j\leq m_{k}}\tau(z_{ki}-z_{kj})\right) \\
\times\prod_{1\leq k<l\leq N-1}\prod_{i=2}^{m_{k}}\prod_{j=2}^{m_{l}}%
\varkappa(z_{ki}-z_{lj})\left(  \prod_{k=1}^{N-1}\prod_{i=1}^{N}\prod
_{j=2}^{m}\phi(z_{kj}-\theta_{i})\right)  r
\end{multline*}
with $r=$
\begin{multline*}
\operatorname*{Res}_{\theta_{N-1N}=i\eta}\cdots\operatorname*{Res}%
_{\theta_{12}=i\eta}\left(  \prod_{k=1}^{N-1}\int_{\mathcal{C}_{\underline
{\theta}}}dz_{k1}\right)  \prod_{k=1}^{N-1}\left(  \prod_{i=1}^{n}\phi
(z_{k1}-\theta_{i})\prod_{2\leq j\leq m}\tau(z_{k1}-z_{kj})\right) \\
\times\prod_{1\leq k<l\leq N-1}\left(  \varkappa(z_{k1}-z_{l1})\prod
_{i=2}^{m_{k}}\varkappa(z_{ki}-z_{l1})\prod_{j=2}^{m_{l}}\varkappa
(z_{k1}-z_{lj})\right)  p_{n\underline{m}}^{{\mathcal{O}}}(\underline{\theta
},\underline{z})\,.
\end{multline*}
Replacing $\mathcal{C}_{\underline{\theta}}$ by $\mathcal{C}_{\underline
{\theta}^{\prime}}$ where $\underline{\theta}^{\prime}=(\theta_{N+1}%
,\dots,\theta_{n})$ we have used $\tau(0)=\tau(\pm i\eta)=\varkappa
(0)=\varkappa(-i\eta)=0$ and the fact that the $z_{k1}$-integrations give
non-vanishing results only for $z_{k1}=\theta_{k}$ and $\theta_{k+1}%
,\ k=1,\dots,N-1$. This is because for $\theta_{12},\dots,\theta
_{N-1N}\rightarrow i\eta$ pinching appears at $\left(  z_{11}\ldots
,,z_{N-11}\right)  =\left(  \theta_{2},\ldots,\theta_{N}\right)  $ and
$\left(  \theta_{1},\ldots,\theta_{N-1}\right)  $. Defining the function%
\begin{multline*}
f(z_{11},\ldots,z_{N-11})=\prod_{k=1}^{N-1}\prod_{2\leq j\leq m_{k}}%
\tau(z_{k1}-z_{kj})\prod_{1\leq k<l\leq N-1}\varkappa(z_{k1}-z_{l1})\\
\times\prod_{1\leq k<l\leq N-1}\left(  \prod_{j=2}^{m_{l}}\varkappa
(z_{k1}-z_{li})\prod_{i=2}^{m_{k}}\varkappa(z_{ki}-z_{l1})\right)
p_{n\underline{m}}^{{\mathcal{O}}}(\underline{\theta},\underline{z})
\end{multline*}
one obtains by means of (\ref{tk}) after some lengthy but straight forward
calculation $r=$
\begin{multline*}
\operatorname*{Res}_{\theta_{N-1N}=i\eta}\cdots\operatorname*{Res}%
_{\theta_{12}=i\eta}\left(  \prod_{k=1}^{N-1}\int_{\mathcal{C}_{\underline
{\theta}}}\frac{dz_{k1}}{R}\right)  \prod_{k=1}^{N-1}\left(  \prod_{i=1}%
^{n}\phi(z_{k1}-\theta_{i})\right)  f(z_{11},\ldots,z_{N-11})\\
=\left(  \operatorname*{Res}_{\theta=i\eta}\phi(-\theta)\prod_{k=1}^{N-2}%
\phi(ki\eta)\right)  ^{N-1}\left(  \prod_{k=1}^{N-1}\prod_{i=N+1}^{n}%
\phi(\theta_{k+1i})\right) \\
\times\left(  f(\theta_{2},\ldots,\theta_{N})-\left(  \prod_{i=N+1}^{n}%
\frac{\phi(\theta_{1i})}{\phi(\theta_{Ni})}\right)  f(\theta_{1},\ldots
,\theta_{N-1})\right)  \,.
\end{multline*}
It has been used that $f(\ldots,z,\ldots,z\ldots)=f(\ldots,z,\ldots
,z-i\eta\ldots)=0$ because of $\varkappa(0)=\varkappa(-i\eta)=0$. Using
further the defining relation of $\phi$ in terms of $F$ (\ref{phiF}), the Jost
property (\ref{J}) of the $\phi$-function and the properties (iii') of
(\ref{p}) for the p-function we get
\begin{multline*}
\operatorname*{Res}_{\theta_{N-1N}=i\eta}\cdots\operatorname*{Res}%
_{\theta_{12}=i\eta}I_{n\underline{m}}(\underline{\theta},p_{n\underline{m}%
}^{{\mathcal{O}}})=\left(  \operatorname*{Res}_{\theta=i\eta}\phi
(-\theta)\right)  ^{N-1}\prod_{k=1}^{N-2}\phi^{k}(ki\eta)\\
\times I_{n-N\underline{m}-1}(\underline{\theta^{\prime}},p_{n-N\underline
{m}-1}^{{\mathcal{O}}})\left(  \prod_{k=1}^{N}\prod_{i=N+1}^{n}F(\theta
_{ki})\right)  ^{-1}\left(  1-\sigma_{1}^{{\mathcal{O}}}\prod_{i=N+1}%
^{n}S(\theta_{Ni})\right)
\end{multline*}
which together with the relation for the normalization constants (\ref{n})
proves the claim.
\end{proof}

\section{ Proof of the commutation rules \label{a3}}

In this Appendix we prove that we find the same commutation rules for two
fields $\phi(x)$ and $\psi(y)$ when there are bound states poles or even when
the S-matrix has double poles\footnote{These poles appear typically for bound
state - bound state scattering. The case of higher order poles may be
discussed similarly and will be published elsewhere.}.

\textbf{Bound states:} We now show that the same result (\ref{5.11}) appears
when there are bound states\footnote{Here we follow the arguments of Quella
\cite{Q}.} which means that there are poles in the physical strip. Let
$\gamma=(\alpha\beta)$ be a bound state of $\alpha$ and $\beta$ with fusion
angle $\eta_{\alpha\beta}^{\gamma}$ which means that at $\theta_{\alpha\beta
}=i\eta_{\alpha\beta}^{\gamma}$ the S-matrix $S_{\alpha\beta}(\theta)$ has a
pole. The momentum and the rapidity of the bound state are
\begin{align*}
p_{\gamma}  &  =p_{\alpha}+p_{\beta}\\
\theta_{\gamma}  &  =\theta_{\alpha}-i(\pi-\eta_{\bar{\gamma}\alpha}%
^{\bar{\beta}})=\theta_{\beta}+i(\pi-\eta_{\beta\bar{\gamma}}^{\bar{\alpha}})
\end{align*}
where $\eta_{\bar{\gamma}\alpha}^{\bar{\beta}}$ and $\eta_{\beta\bar{\gamma}%
}^{\bar{\alpha}}$ are the fusion angles of the bound states $\bar{\beta}%
=(\bar{\gamma}\alpha)$ and $\bar{\alpha}=(\beta\bar{\gamma})$, respectively.

We start matrix element of $\psi(y)\phi(x)$ (given by (\ref{phipsi}) with
(\ref{Y})). First we consider the contribution in the sum over the
intermediate states where $\alpha\in\underline{\bar{\omega}}$ and $\beta
\in\underline{\lambda}$. All the particles which are not essential for this
discussion will be suppressed. Then the function $\psi_{\alpha\beta}%
(\tilde{\theta}_{\alpha},\theta_{\beta})$ has a pole at $\tilde{\theta
}_{\alpha}-\theta_{\beta}=i\eta_{\alpha\beta}^{\gamma}$ such that by shifting
the integration $\tilde{\theta}_{\alpha}\rightarrow\tilde{\theta}_{\alpha
}+i\pi_{-}$ there will be the additional contribution
\begin{align*}
&  \frac{i}{2}\operatorname*{Res}_{\tilde{\theta}_{\alpha}=\theta_{\alpha}%
}\phi_{\bar{\alpha}}(\tilde{\theta}_{\alpha}-i\pi_{-})\mathbf{C}^{\bar{\alpha
}\alpha}\psi_{\alpha\beta}(\tilde{\theta}_{\alpha},\theta_{\beta}%
)e^{i\tilde{P}_{\alpha}(x-y)}e^{-iyP_{\beta}}\\
&  =\frac{i}{2}\phi_{\bar{\alpha}}(\theta_{\alpha}-i\pi_{-})\mathbf{C}%
^{\bar{\alpha}\alpha}\psi_{\gamma}(\theta_{\gamma})\sqrt{2}\Gamma_{\alpha
\beta}^{\gamma}e^{ixP_{\alpha}-iyP_{\gamma}}%
\end{align*}
with $\theta_{\alpha}=\theta_{\beta}+i\eta_{\alpha\beta}^{\gamma}%
,\;\theta_{\gamma}=\theta_{\beta}+i(\pi-\eta_{\beta\gamma}^{\bar{\alpha}})$
and the fusion intertwiner $\Gamma_{\alpha\beta}^{\gamma}$ (see e.g.
\cite{BK1}). Next we consider the contribution to the sum over the
intermediate states where $\gamma\in\underline{\bar{\omega}}$ and$\;\beta
=\underline{\mu}$. Then the function $\phi_{\beta\bar{\gamma}}(\theta_{\beta
},\tilde{\theta}_{\gamma}-i\pi)$ has a pole at $\theta_{\beta}-\tilde{\theta
}_{\gamma}+i\pi=i\eta_{\beta\bar{\gamma}}^{\bar{\alpha}}$ such that by
shifting the integration $\tilde{\theta}_{\gamma}\rightarrow\tilde{\theta
}_{\gamma}+i\pi_{-}$ there will be the additional contribution
\begin{align*}
&  \frac{i}{2}\operatorname*{Res}_{\tilde{\theta}_{\gamma}=\theta_{\gamma}%
}\phi_{\beta\bar{\gamma}}(\theta_{\beta},\tilde{\theta}_{\gamma}-i\pi
_{-})\mathbf{C}^{\bar{\gamma}\gamma}\psi_{\gamma}(\tilde{\theta}_{\gamma
})e^{i\tilde{P}_{\gamma}(x-y)}e^{-ixP_{\beta}}\\
&  =-\frac{i}{2}\phi_{\bar{\alpha}}(\theta_{\alpha}-i\pi_{-})\sqrt{2}%
\Gamma_{\beta\bar{\gamma}}^{\bar{\alpha}}\mathbf{C}^{\bar{\gamma}\gamma}%
\psi_{\gamma}(\theta_{\gamma})e^{ixP_{\alpha}-iyP_{\gamma}}%
\end{align*}
with $\theta_{\alpha},\theta_{\gamma}$ as above and the fusion intertwiner
$\Gamma_{\beta\gamma}^{\bar{\alpha}}$. From the crossing relation of the
fusion intertwiners%
\[
\mathbf{C}^{\bar{\alpha}\alpha}\Gamma_{\alpha\beta}^{\gamma}=\Gamma_{\beta
\bar{\gamma}}^{\bar{\alpha}}\mathbf{C}^{\bar{\gamma}\gamma}%
\]
we conclude that these residue terms form bound state poles cancel. The steps
of the argument may be depicted as%
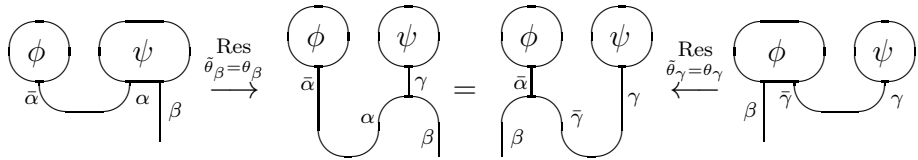
\[%
\begin{array}
[c]{c}%
\mbox{
\unitlength4mm\begin{picture}(6,4)
\put(1,3){\oval(2,2)}
\put(.6,2.7){$\phi$}
\put(4.5,3){\oval(3,2)}
\put(4.1,2.7){$\psi$}
\put(5,0){\line(0,1){2}}
\put(2.5,2){\oval(3,2)[b]}
\put(.5,1.4){$_{\bar\alpha}$}
\put(4.2,1.4){$_\alpha$}
\put(5.3,1){$_\beta$}
\end{picture}}
\end{array}
\overset{\operatorname*{Res}\limits_{\tilde{\theta}_{\beta}=\theta_{\beta}}%
}{\longrightarrow}%
\begin{array}
[c]{c}%
\mbox{
\unitlength4mm\begin{picture}(5,5)
\put(1,4){\oval(2,2)}
\put(.6,3.7){$\phi$}
\put(1,1){\line(0,1){2}}
\put(4,4){\oval(2,2)}
\put(3.6,3.7){$\psi$}
\put(2,1){\oval(2,2)[b]}
\put(4,1){\oval(2,2)[t]}
\put(4,2){\line(0,1){1}}
\put(5,0){\line(0,1){1}}
\put(4.2,2.5){$_\gamma$}
\put(.4,2.4){$_{\bar\alpha}$}
\put(2.4,1.3){$_\alpha$}
\put(4.4,.6){$_\beta$}
\end{picture}}
\end{array}
=%
\begin{array}
[c]{c}%
\mbox{
\unitlength4mm\begin{picture}(5,5)
\put(1,4){\oval(2,2)}
\put(.6,3.7){$\phi$}
\put(4,4){\oval(2,2)}
\put(3.6,3.7){$\psi$}
\put(4,1){\line(0,1){2}}
\put(3,1){\oval(2,2)[b]}
\put(1,1){\oval(2,2)[t]}
\put(1,2){\line(0,1){1}}
\put(0,0){\line(0,1){1}}
\put(4.2,2){$_\gamma$}
\put(.4,2.4){$_{\bar\alpha}$}
\put(2.3,1.3){$_{\bar \gamma}$}
\put(.3,.6){$_\beta$}
\end{picture}}
\end{array}
\overset{\operatorname*{Res}\limits_{\tilde{\theta}_{\gamma}=\theta_{\gamma}}%
}{\longleftarrow}%
\begin{array}
[c]{c}%
\mbox{
\unitlength4mm\begin{picture}(6,4)(.5,0)
\put(1.5,3){\oval(3,2)}
\put(1.1,2.7){$\phi$}
\put(5,3){\oval(2,2)}
\put(4.6,2.7){$\psi$}
\put(1,0){\line(0,1){2}}
\put(3.5,2){\oval(3,2)[b]}
\put(.3,1){$_{\beta}$}
\put(1.5,1.4){$_{\bar\gamma}$}
\put(5.2,1.4){$_\gamma$}
\end{picture}}
\end{array}
\]

\textbf{Double poles:\mbox{\protect\footnotemark}} \footnotetext{This
discussion is new.} Form factors have more poles which are not related to
bound states, they belong to higher poles of the S-matrix. First we consider
the contribution to the sum over the intermediate states where the particles
$\bar{1},1\in\underline{\bar{\omega}}$,$\;2\in\underline{\mu}$. Again we
suppress all particles which are not essential for our discussion. Then the
function $\phi_{2\bar{1}1}(\theta,\tilde{\theta}-i\pi,\tilde{\theta}^{\prime
}-i\pi)$ has a pole at $\tilde{\theta}=\theta_{1}=\theta+i\pi/N$ which
correspond to the bound state $(2\bar{1})=1$ with the fusion angle
$\eta_{2\bar{1}}^{1}=\pi(1-1/N)$. We shift the integrations $\tilde{\theta
}\rightarrow\tilde{\theta}+i\pi_{-}$ and $\tilde{\theta}^{\prime}%
\rightarrow\tilde{\theta}^{\prime}+i\pi_{-}$ such that during the shift
$0<\operatorname{Im}(\tilde{\theta}-\tilde{\theta}^{\prime})<\epsilon$. From
the $\tilde{\theta}$-integration there will be the additional contribution
\begin{align*}
&  \frac{i}{2}\operatorname*{Res}_{\tilde{\theta}=\theta_{1}}\phi_{2\bar{1}%
1}(\theta,\tilde{\theta}-i\pi,\tilde{\theta}^{\prime}-i\pi)\mathbf{C}^{\bar
{1}11\bar{1}}\psi_{1\bar{1}}(\tilde{\theta}^{\prime},\tilde{\theta}%
)e^{i\tilde{P}(x-y)}e^{-ixP}\\
&  =-\frac{i}{2}\phi_{11}(\theta-i\pi/N,\tilde{\theta}^{\prime}-i\pi)\sqrt
{2}\Gamma_{2}^{11}\mathbf{C}^{\bar{1}11\bar{1}}\psi_{1\bar{1}}(\tilde{\theta
}^{\prime},\theta_{1})e^{i\tilde{P}(x-y)}e^{-ixP}\,.
\end{align*}
Further the function $\phi_{11}(\theta-i\pi/N,\tilde{\theta}^{\prime}-i\pi)$
has a pole at $\tilde{\theta}^{\prime}=\theta_{2}=\theta+i\pi(1-3/N)$ which
correspond to the bound state $(11)=2$ with the fusion angle $\eta_{11}%
^{2}=\pi2/N$. From the $\tilde{\theta}^{\prime}$-integration there will be the
additional contribution
\begin{align*}
&  \left(  -\frac{i}{2}\right)  ^{2}\sqrt{2}\operatorname*{Res}_{\tilde
{\theta}^{\prime}=\theta_{2}}\phi_{11}(\theta-i\pi/N,\tilde{\theta}^{\prime
}-i\pi)\sqrt{2}\Gamma_{2}^{11}\mathbf{C}^{\bar{1}11\bar{1}}\psi_{1\bar{1}%
}(\tilde{\theta}^{\prime},\theta_{1})e^{i\tilde{P}(x-y)}e^{-ixP}\\
&  =-\frac{1}{2}\phi_{2}(\theta-i\pi2/N)\Gamma_{11}^{2}\Gamma_{2}%
^{11}\mathbf{C}^{\bar{1}11\bar{1}}\psi_{1\bar{1}}(\theta_{2},\theta
_{1})e^{i(P_{1}+P_{2})(x-y)}e^{-ixP}%
\end{align*}
This procedure may be depicted as
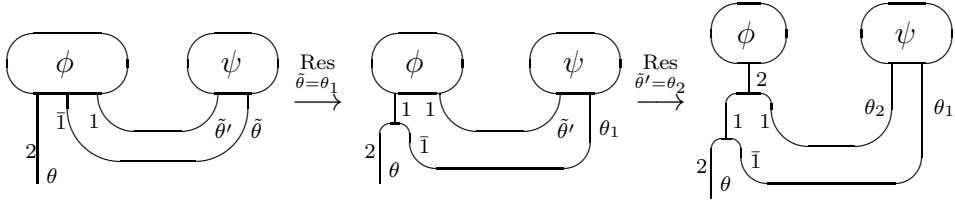
\[%
\begin{array}
[c]{c}%
\mbox{
\unitlength4mm\begin{picture}(9,6)
\put(2,4){\oval(4,2)}
\put(1.6,3.7){$\phi$}
\put(1,0){\line(0,1){3}}
\put(1.3,.2){$_\theta$}
\put(.6,1){$_2$}
\put (2.7,2){$_1$}
\put(1.6,2){$_{\bar1}$}
\put(5,3){\oval(4,2.5)[b]}
\put (5,3){\oval(6,4.5)[b]}
\put(6.9,1.8){$_{\tilde\theta'}$}
\put(8.1,1.8){$_{\tilde \theta}$}
\put(7.5,4){\oval(3,2)}
\put(7.1,3.7){$\psi$} \end{picture}}
\end{array}
\overset{\operatorname*{Res}\limits_{\tilde{\theta}=\theta_{1}}}%
{\longrightarrow}%
\begin{array}
[c]{c}%
\mbox{
\unitlength4mm\begin{picture}(8.4,6)(.6,0)
\put(2.25,4){\oval(3,2)}
\put(1.85,3.7){$\phi$}
\put(1,0){\line(0,1){1.5}}
\put(1.5,2){\line(0,1){1}}
\put(8,1.5){\line (0,1){1.5}}
\put(1.3,.4){$_\theta$}
\put(.5,1){$_2$}
\put(2.5,2.4){$_1$}
\put (1.7,2.4){$_1$}
\put(2.3,1.2){$_{\bar1}$}
\put(5,3){\oval(4,2.5)[b]}
\put (5,1.5){\oval(6,2)[b]}
\put(1.5,1.5){\oval(1,1)[t]}
\put(6.9,1.8){$_{\tilde \theta'}$}
\put(8.3,1.8){$_{\theta_1}$}
\put(7.5,4){\oval(3,2)}
\put (7.1,3.7){$\psi$}\end{picture}}
\end{array}
\overset{\operatorname*{Res}\limits_{\tilde{\theta}^{\prime}=\theta_{2}}%
}{\longrightarrow}%
\begin{array}
[c]{c}%
\mbox{
\unitlength4mm\begin{picture}(8,7)(1,0)
\put(2.25,5.5){\oval(2.5,2)}
\put (1.85,5.2){$\phi$}
\put(1,0){\line(0,1){1.5}}
\put(1.5,2){\line(0,1){1}}
\put(7,3){\line(0,1){1.5}}
\put(8,1.5){\line(0,1){3}}
\put(2.25,3.5){\line (0,1){1}}
\put(1.3,.4){$_\theta$}
\put(.5,1){$_2$}
\put(2.5,3.9){$_2$}
\put (2.6,2.5){$_1$}
\put(1.7,2.5){$_1$}
\put(2.3,1.2){$_{\bar1}$}
\put(5,3){\oval (4,2.5)[b]}
\put(5,1.5){\oval(6,2)[b]}
\put(1.5,1.5){\oval(1,1)[t]}
\put(2.25,3){\oval(1.5,1)[t]}
\put(6.1,3){$_{\theta_2}$}
\put(8.4,3){$_{\theta _1}$}
\put(7.5,5.5){\oval(3,2)}
\put(7.1,5.2){$\psi$} \end{picture}}
\end{array}
\]
We show this the additional term is cancelled by a contribution to the sum
over the intermediate states where $\bar{2}\in\underline{\omega}%
,\ 2\in\underline{\lambda}$. Then the function $\psi_{\bar{2}2}(\tilde{\theta
},\theta)$ has a pole at $\tilde{\theta}=\theta_{3}=\theta+i\pi(1-2/N)$ which
correspond to the double pole of the S-matrix
\[
S_{\bar{2}2}(\theta)=\left(  \frac{\sin\frac{\pi}{2}\left(  \frac{\theta}%
{i\pi}+\frac{N-2}{N}\right)  }{\sin\frac{\pi}{2}\left(  \frac{\theta}{i\pi
}-\frac{N-2}{N}\right)  }\right)  ^{2}\frac{\sin\frac{\pi}{2}\left(
\frac{\theta}{i\pi}+\frac{N-4}{N}\right)  }{\sin\frac{\pi}{2}\left(
\frac{\theta}{i\pi}-\frac{N-4}{N}\right)  }%
\]
at $\theta=i\pi(1-2/N)$. From the $\tilde{\theta}$-integration there will be
the additional contribution
\begin{align*}
&  \frac{i}{2}\operatorname*{Res}_{\tilde{\theta}=\theta_{3}}\phi_{2}%
(\tilde{\theta}-i\pi)\mathbf{C}^{2\bar{2}}\psi_{\bar{2}2}(\tilde{\theta
},\theta)e^{i\tilde{P}(x-y)}e^{-iyP}\\
&  =\frac{i}{2}i\phi_{2}(\theta-i\pi2/N)\mathbf{C}^{2\bar{2}}(-i)\left(
\psi_{\bar{1}1}(\theta_{2},\theta_{1})\Gamma_{\bar{2}}^{\bar{1}\bar{1}%
}\mathbf{C}_{\bar{1}1}\Gamma_{2}^{11}\right)  e^{iP_{3}(x-y)}e^{-iyP}\,.
\end{align*}
This procedure may be depicted as%
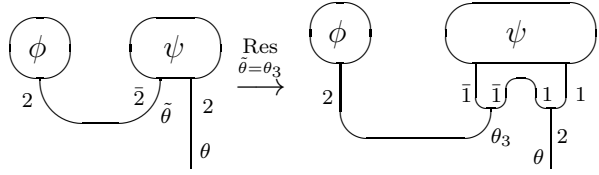
\[%
\begin{array}
[c]{c}%
\mbox{
\unitlength4mm\begin{picture}(7,6)
\put(1,4){\oval(2,2)}
\put(.6,3.7){$\phi$}
\put(3,3){\oval(4,3)[b]}
\put(.4,2.2){$_2$}
\put(4.1,2.3){$_{\bar2}$}
\put(6.4,2){$_2$}
\put(5,1.8){$_{\tilde\theta}$}
\put(6.3,.5){$_\theta$}
\put(6,0){\line(0,1){3}}
\put(5.5,4){\oval(3,2)}
\put(5.1,3.7){$\psi$}
\end{picture}}
\end{array}
\overset{\operatorname*{Res}\limits_{\tilde{\theta}=\theta_{3}}}%
{\longrightarrow}%
\begin{array}
[c]{c}%
\mbox{
\unitlength4mm\begin{picture}(9,6)
\put(1,4.5){\oval(2,2)}
\put(.6,4.2){$\phi$}
\put(3.5,2){\oval(5,2)[b]}
\put(8.8,2.4){$_1$}
\put(7.7,2.4){$_1$}
\put(6,2.4){$_{\bar1}$}
\put(5,2.4){$_{\bar1}$}
\put(.4,2.3){$_2$}
\put(8.2,1){$_2$}
\put(6,1){$_{\theta_3}$}
\put(7.4,.3){$_\theta$}
\put(1,2){\line(0,1){1.5}}
\put(5.5,2.5){\line(0,1){1}}
\put(8.5,2.5){\line(0,1){1}}
\put(6,2.5){\oval(1,1)[b]}
\put(7,2.5){\oval(1,1)[t]}
\put(8,2.5){\oval(1,1)[b]}
\put(7,4.5){\oval(5,2)}
\put(6.5,4.2){$\psi$}
\put(8,0){\line(0,1){2}}
\end{picture}}
\end{array}
\]
The crossing relation of the fusion intertwiners
\[
\Gamma_{11}^{2}\Gamma_{2}^{11}\mathbf{C}^{1\bar{1}}=\mathbf{C}^{2\bar{2}%
}\Gamma_{\bar{2}}^{\bar{1}\bar{1}}\mathbf{C}_{\bar{1}1}\Gamma_{2}^{11}%
\]
implies that this contribution again cancelled the one above. It has been used
that the form factor of bound states $\bar{2}2$ has a simple pole where the
S-matrix $S_{\bar{2}2}$ has a double pole and the residue is%
\[
\operatorname*{Res}_{\tilde{\theta}=\theta_{3}}\psi_{\bar{2}2}(\tilde{\theta
},\theta)=-i\left(  \psi_{\bar{1}1}(\theta_{2},\theta_{3})\Gamma_{\bar{2}%
}^{\bar{1}\bar{1}}\mathbf{C}_{\bar{1}1}\Gamma_{2}^{11}\right)  \,.
\]
This may be calculated as follows. By the form factor equation (iv) we have%
\[
\operatorname*{Res}_{\theta_{12}=i\eta_{11}^{2}}\operatorname*{Res}%
_{\theta_{34}=i\eta_{11}^{2}}\psi_{\bar{1}\bar{1}11}(\theta_{1},\theta
_{2},\theta_{3},\theta_{4})=2\psi_{\bar{2}2}(\theta_{(12)},\theta
_{(34)})\Gamma_{\bar{1}\bar{1}}^{\bar{2}}\Gamma_{11}^{2}\,.
\]
Therefore using the form factor equation (iii) and the definition of the
fusion intertwiners $\operatorname*{Res}iS=\Gamma\Gamma$ we obtain%
\begin{align*}
&  \operatorname*{Res}_{\theta_{(12)(34)}=i\pi(1-2/N)}\psi_{\bar{2}2}%
(\theta_{(12)},\theta_{(34)})\Gamma_{\bar{1}\bar{1}}^{\bar{2}}\Gamma_{11}%
^{2}\\
&  =\frac{1}{2}\operatorname*{Res}_{\theta_{14}=i\pi}\operatorname*{Res}%
_{\theta_{12}=i\eta_{11}^{2}}\operatorname*{Res}_{\theta_{34}=i\eta_{11}^{2}%
}\psi_{\bar{1}\bar{1}11}(\theta_{2},\theta_{1},\theta_{4},\theta_{3}%
)S_{\bar{1}\bar{1}}(\theta_{12})S_{11}(\theta_{34})\\
&  =\frac{1}{2}\operatorname*{Res}_{\theta_{14}=i\pi}\operatorname*{Res}%
_{\theta_{12}=i\eta_{11}^{2}}2i\mathbf{C}_{\bar{1}1}\left(  \psi_{\bar{1}%
1}(\theta_{2},\theta_{3})S_{\bar{1}\bar{1}}(\theta_{12})S_{11}(\theta
_{34})-\psi_{\bar{1}1}(\theta_{2},\theta_{3})\right) \\
&  =i\mathbf{C}_{\bar{1}1}\left(  \psi_{\bar{1}1}(\theta_{2},\theta
_{3})(-i)\Gamma_{\bar{2}}^{\bar{1}\bar{1}}\Gamma_{\bar{1}\bar{1}}^{\bar{2}%
}(-i)\Gamma_{2}^{11}\Gamma_{11}^{2}\right)
\end{align*}
which implies the residue formula used above. This procedure may be depicted
as%
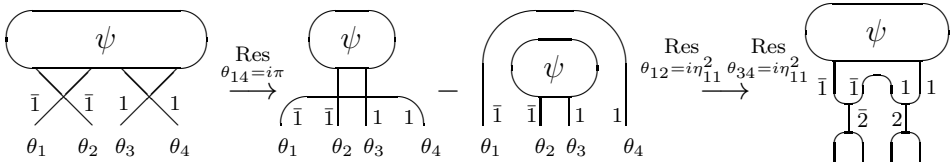
\[%
\begin{array}
[c]{c}%
\mbox{
\unitlength3.8mm\begin{picture}(7,5)
\put(3.5,4){\oval(7,2)}
\put(3.1,3.7){$\psi$}
\put(1,1){\line(1,1){2}}
\put(3,1){\line(-1,1){2}}
\put(4,1){\line(1,1){2}}
\put(6,1){\line(-1,1){2}}
\put(.7,.2){$_{\theta_1}$}
\put(2.5,.2){$_{\theta_2}$}
\put(3.8,.2){$_{\theta_3}$}
\put(5.7,.2){$_{\theta_4}$}
\put(.8,1.7){$_{\bar1}$}
\put(2.7,1.7){$_{\bar1}$}
\put(3.9,1.7){$_1$}
\put(5.7,1.7){$_1$}
\end{picture}}
\end{array}
\overset{\operatorname*{Res}\limits_{\theta_{14}=i\pi}}{\longrightarrow}%
\begin{array}
[c]{c}%
\mbox{
\unitlength3.8mm\begin{picture}(4,5)(1,0)
\put(2.5,4){\oval(3,2)}
\put(2.1,3.7){$\psi$}
\put(2,1){\line(0,1){2}}
\put(3,1){\line(0,1){2}}
\put(2.5,1){\oval(5,2)[t]}
\put(-.1,.2){$_{\theta_1}$}
\put(1.8,.2){$_{\theta_2}$}
\put(2.9,.2){$_{\theta_3}$}
\put(4.9,.2){$_{\theta_4}$}
\put(.4,1.2){$_{\bar1}$}
\put(1.5,1.2){$_{\bar1}$}
\put(3.2,1.2){$_1$}
\put(4.3,1.2){$_1$}
\end{picture}}
\end{array}
-%
\begin{array}
[c]{c}%
\mbox{
\unitlength3.8mm\begin{picture}(5,5)
\put(2.5,3){\oval(3,2)}
\put(2.1,2.7){$\psi$}
\put(2,1){\line(0,1){1}}
\put(3,1){\line(0,1){1}}
\put(2.5,1){\oval(5,8)[t]}
\put(-.1,.2){$_{\theta_1}$}
\put(1.8,.2){$_{\theta_2}$}
\put(2.9,.2){$_{\theta_3}$}
\put(4.9,.2){$_{\theta_4}$}
\put(.3,1.3){$_{\bar1}$}
\put(1.5,1.3){$_{\bar1}$}
\put(3.2,1.3){$_1$}
\put(4.5,1.3){$_1$}
\end{picture}}
\end{array}
\overset{\operatorname*{Res}\limits_{\theta_{12}=i\eta_{11}^{2}}%
\operatorname*{Res}\limits_{\theta_{34}=i\eta_{11}^{2}}}{\longrightarrow}%
\begin{array}
[c]{c}%
\mbox{
\unitlength3.8mm\begin{picture}(4,5.5)(1,0)
\put(2.5,4.5){\oval(5,2)}
\put(2.1,4.2){$\psi$}
\put(.4,2.5){$_{\bar1}$}
\put(1.5,2.5){$_{\bar1}$}
\put(3.3,2.5){$_1$}
\put(4.2,2.5){$_1$}
\put(1.8,1.4){$_{\bar2}$}
\put(3,1.4){$_2$}
\put(1.5,2.5){\oval(1,1)[b]}
\put(2.5,2.5){\oval(1,1)[t]}
\put(3.5,2.5){\oval(1,1)[b]}
\put(1,2.5){\line(0,1){1}}
\put(4,2.5){\line(0,1){1}}
\put(1.5,1){\line(0,1){1}}
\put(3.5,1){\line(0,1){1}}
\put(1.5,0){\oval(1,2)[t]}
\put(3.5,0){\oval(1,2)[t]}
\end{picture}}
\end{array}
\]
Note that the last graph, as an on-shell graph, resembles (half of) the `box'
Feynman diagram which was used to investigate the double poles of bound state
S-matrices (see e.g. \cite{CT,BCDS})

\textbf{The general case:} Finally we consider the general case that the sets
rapidities in the initial state $\underline{\theta}_{\alpha}$ and the ones
final state $\underline{\theta}_{\beta}^{\prime}$ have also common elements.
Then after inserting (\ref{5.93}) and (\ref{5.95}) in (\ref{5.9}) there will
be S-matrices $S_{\underline{\nu}\underline{\pi}}^{\underline{\varsigma
}\underline{\sigma}}(\underline{\tilde{\theta}}_{\nu},\underline{\tilde
{\theta}}_{\pi})$ which produce additional poles in the physical strip which
would produce additional residue contributions while shifting the integration
contours. However, we can remove these S-matrices by using again the crossing
relation (ii) and move all the lines of common rapidities to the left or right
as depicted as follows%
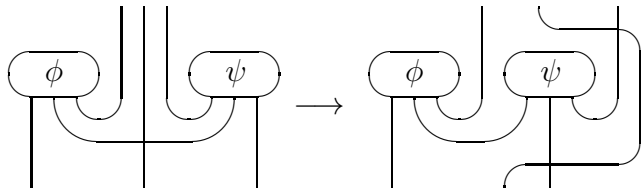
\[%
\begin{array}
[c]{c}%
\unitlength3mm
\begin{picture}(12,8) \put(2,5){\oval(4,2)} \put(1.6,4.7){$\phi$}
\put(1,0){\line(0,1){4}} \put(4,4){\oval(2,2)[b]} \put(5,4){\line(0,1){4}}
\put(6,4){\oval(8,4)[b]} \put(10,5){\oval(4,2)} \put(9.6,4.7){$\psi$}
\put(8,4){\oval(2,2)[b]} \put(7,4){\line(0,1){4}} \put(6,0){\line(0,1){8}}
\put(11,0){\line(0,1){4}} \end{picture}
\end{array}
\longrightarrow%
\begin{array}
[c]{c}%
\mbox{
\unitlength3mm\begin{picture}(12,8)
\put(2,5){\oval(4,2)}
\put(1.6,4.7){$\phi$}
\put(1,0){\line(0,1){4}}
\put(4,4){\oval(2,2)[b]}
\put(5,4){\line(0,1){4}}
\put(4.5,4){\oval(5,4)[b]}
\put(8,5){\oval(4,2)}
\put(7.6,4.7){$\psi$}
\put(10,4){\oval(2,2)[b]}
\put(11,4){\line(0,1){4}}
\put(10,8){\oval(5,2)[bl]}
\put(10,6){\oval(4,2)[tr]}
\put(12,2){\line(0,1){4}}
\put(9,0){\oval(6,2)[tl]}
\put(9,2){\oval(6,2)[br]}
\put(8,0){\line(0,1){4}}
\end{picture}}
\end{array}
\]
Then we can apply the procedure as above.


\begin{thebibliography}{10}

\bibitem{KW}
M.~Karowski and P.~Weisz,
\newblock Nucl. Phys. {\bf B139} (1978) 455.

\bibitem{BFKZ}
H.~M. Babujian, A.~Fring, M.~Karowski, and A.~Zapletal,
\newblock Nucl. Phys. {\bf B538} (1999) 535.

\bibitem{Sm}
F.~Smirnov,
\newblock Adv. Series in Math. Phys. \textbf{14}, World Scientific, 1992.

\bibitem{BK}
H.~Babujian and M.~Karowski,
\newblock Nucl. Phys. {\bf B620} (2002) 407.

\bibitem{K1}
M.~Karowski,
\newblock Nucl. Phys. {\bf B153} (1979) 244.

\bibitem{Wa}
K.~M. Watson,
\newblock Phys. Rev. {\bf 95} (1954) 228.

\bibitem{B}
H.~M. Babujian,
\newblock In `Gosen 1990, Proceedings, Theory of elementary particles' 12-23.
  (see high energy physics index 29 (1991) No. 12257).

\bibitem{B1}
H.~Babujian,
\newblock J. Phys. {\bf A26} (1993) 6981.

\bibitem{KS}
R.~K{\"o}berle and J.~A. Swieca,
\newblock Phys. Lett. {\bf B86} (1979) 209.

\bibitem{Bariev}
R.~Z. Bariev,
\newblock Phys. Lett. {\bf A55} (1976) 456.

\bibitem{MTW}
B.~M. McCoy, C.~A. Tracy, and T.~T. Wu,
\newblock Phys. Rev. Lett. {\bf 38} (1977) 793.

\bibitem{SJM1}
M.~Sato, T.~Miwa, and M.~Jimbo,
\newblock Proc. Japan Acad. {\bf 53} (1977) 6.

\bibitem{BKW}
B.~Berg, M.~Karowski, and P.~Weisz,
\newblock Phys. Rev. {\bf D19} (1979) 2477.

\bibitem{Za}
A.~B. Zamolodchikov,
\newblock Int. J. Mod. Phys. {\bf A3} (1988) 743.

\bibitem{Fa}
V.~A. Fateev,
\newblock Int. J. Mod. Phys. {\bf A6} (1991) 2109.

\bibitem{FZ}
V.~A. Fateev and A.~B. Zamolodchikov,
\newblock Sov. Phys. JETP {\bf 62} (1985) 215.

\bibitem{K3}
M.~Karowski,
\newblock Lecture Notes in Physics (Springer) {\bf 126} (1979) 344.

\bibitem{KiSm}
A.~N. Kirillov and F.~A. Smirnov,
\newblock ITF preprint 88-73P, Kiev  (1988).

\bibitem{DC}
G.~Delfino and J.~L. Cardy,
\newblock Nucl. Phys. {\bf B519} (1998) 551.

\bibitem{JKOPS}
M.~Jimbo, H.~Konno, S.~Odake, Y.~Pugai, and J.~Shiraishi,
\newblock J. Statist. Phys. {\bf 102} (2001) 883.

\bibitem{EK}
F.~H.~L. Essler and R.~M. Konik,
\newblock (2004),
\newblock In `Shifman, M. (ed.) et al.: From fields to strings, vol. 1'
  684-830.

\bibitem{Ts}
A.~M. Tsvelik,
\newblock Cambridge, UK: Univ. Pr. (1995) 332 p.

\bibitem{GNT}
A.~O. Gogolin, A.~A. Nersesyan, and A.~M. Tsvelik,
\newblock Cambridge, UK: Univ. Pr. (1999).

\bibitem{LZMGo}
J.~Links, H.~Zhou, R.~McKenzie, and M.~Gould,
\newblock J. Phys. {\bf A36} (2003) R63.

\bibitem{BK04}
H.~Babujian and M.~Karowski,
\newblock Phys. Lett. {\bf B575} (2003) 144.

\bibitem{BK2}
H.~Babujian and M.~Karowski,
\newblock J. Phys. {\bf A35} (2002) 9081.

\bibitem{Las}
M.~Y. Lashkevich,
\newblock (1994),
\newblock LANDAU-94-TMP-4, unpublished.

\bibitem{Q}
T.~Quella,
\newblock (1999),
\newblock Diploma thesis FU-Berlin (1999), unpublished.

\bibitem{CT}
S.~R. Coleman and H.~J. Thun,
\newblock Commun. Math. Phys. {\bf 61} (1978) 31.

\bibitem{BCDS}
H.~W. Braden, E.~Corrigan, P.~E. Dorey, and R.~Sasaki,
\newblock Nucl. Phys. {\bf B338} (1990) 689.

\bibitem{DM}
G.~Delfino and G.~Mussardo,
\newblock Nucl. Phys. {\bf B455} (1995) 724.

\bibitem{AMV}
C.~Acerbi, G.~Mussardo, and A.~Valleriani,
\newblock J. Phys. {\bf A30} (1997) 2895.

\bibitem{BK1}
H.~Babujian and M.~Karowski,
\newblock Phys. Lett. {\bf B471} (1999) 53.

\end{thebibliography}

\end{document}